\documentclass[twocolumn,tighten]{aastex62}
\RequirePackage{rotating}
\usepackage{multirow}
\usepackage{graphicx}
\usepackage{grffile}
\usepackage{subfigure}
\usepackage{amsmath}

\usepackage[normalem]{ulem}

\graphicspath{{./}{figures/}}

\usepackage{soul} 

\submitjournal{ApJ}

\shorttitle{Magnetic reconnection in the upper chromosphere}
\shortauthors{Wargnier et al.}

\usepackage{natbib}
\begin{document}

\title{Multi-Fluid Simulations of Upper Chromospheric Magnetic Reconnection with Helium-Hydrogen mixture}

\correspondingauthor{Q. M. Wargnier}
\email{}

\author{Q. M. Wargnier}
\affil{Lockheed Martin Solar \& Astrophysics Laboratory,
3251 Hanover St, Palo Alto, CA 94304, USA}
\affil{Bay Area Environmental Research Institute,
NASA Research Park, Moffett Field, CA 94035, USA}

\author{J. Mart\'inez-Sykora}
\affil{Lockheed Martin Solar \& Astrophysics Laboratory,
3251 Hanover St, Palo Alto, CA 94304, USA}
\affil{Bay Area Environmental Research Institute,
NASA Research Park, Moffett Field, CA 94035, USA}
\affil{Rosseland Centre for Solar Physics, University of Oslo,
P.O. Box 1029 Blindern, NO0315 Oslo, Norway}

\author{V. H. Hansteen}
\affil{Lockheed Martin Solar \& Astrophysics Laboratory,
3251 Hanover St, Palo Alto, CA 94304, USA}
\affil{Bay Area Environmental Research Institute,
NASA Research Park, Moffett Field, CA 94035, USA}
\affil{Rosseland Centre for Solar Physics, University of Oslo,
P.O. Box 1029 Blindern, NO0315 Oslo, Norway}
\affil{Institute of Theoretical Astrophysics, University of Oslo, P.O. Box 1029 Blindern, N-0315 Oslo, Norway}

\author{B. De Pontieu}
\affil{Lockheed Martin Solar \& Astrophysics Laboratory,
3251 Hanover St, Palo Alto, CA 94304, USA}
\affil{Rosseland Centre for Solar Physics, University of Oslo,
P.O. Box 1029 Blindern, NO0315 Oslo, Norway}
\affil{Institute of Theoretical Astrophysics, University of Oslo, P.O. Box 1029 Blindern, N-0315 Oslo, Norway}

\newcommand{\eg}{{\it e.g.,}} 
\newcommand{\ie}{{\it i.e.,}} 
\newcommand{\myemail}{wargnier@baeri.org}
\newcommand{\komment}[1]{\texttt{#1}}
\newcommand{\pref}{\protect\ref}
\newcommand{\soho}{{\em SOHO{}}}
\newcommand{\sdo}{{\em SDO{}}}
\newcommand{\stereo}{{\em STEREO{}}}
\newcommand{\iris}{{\em IRIS{}}}
\newcommand{\hinode}{{\em Hinode{}}}
\newcommand{\ebysus}{{\em Ebysus{}}}
\newcommand{\bifrost}{{\em Bifrost{}}}
\newcommand{\wc}{multi-ion cyclotron}
\newcommand{\hp}{H$^{+}$}
\newcommand{\hep}{He$^{+}$}
\newcommand{\cp}{C$^{+}$}
\newcommand{\hepp}{He$^{++}$}
\newcommand{\jms}[1]{\color{red}{#1}}
\newcommand{\ms}[1]{\color{brown}{#1}}
\newcommand\msrm{\bgroup\markoverwith{\textcolor{magenta}{\rule[0.5ex]{2pt}{0.4pt}}}\ULon}
\newcommand{\velocity}{\bold u}
\newcommand{\thermalspeed}{\bold u^{\text{th}}} 
\newcommand{\viscous}{\boldsymbol{\tau}} 
\newcommand{\identity}{\mathbf{\mathbb{I}}} 
\newcommand{\EField}{\bold E} 
\newcommand{\BField}{\bold B}
\newcommand{\pvec}{\wedge}
\newcommand{\momex}{\bold R}
\newcommand{\dt}{\partial_t}
\newcommand{\totalcurrent}{\bold J}
\newcommand{\mixture}{\mathcal{M}}
\newcommand{\heavy}{\mathcal{H}}
\newcommand{\elec}{\mathfrak{e}} 
\newcommand{\ions}{\mathcal{I}} 
\newcommand{\Q}{\mathfrak{Q}}
\newcommand{\g}{\mathfrak{g}} 
\newcommand{\ionized}{\mathcal{\hat{I}}} 
\newcommand{\specie}{\mathfrak{a}} 
\newcommand{\fact}[1]{#1\mathpunct{}!}
\newcommand{\crosssection}{C}
\newcommand{\refeq}[1]{Eq.~\ref{#1}}
\newcommand{\refsec}[1]{Section~\ref{#1}}
\newcommand{\reffig}[1]{Fig.~\ref{#1}}
\newcommand{\angstrom}{\mbox{\normalfont\AA}}
\newcommand{\overbar}[1]{\mkern 1.5mu\overline{\mkern-1.5mu#1\mkern-1.5mu}\mkern 1.5mu}

\begin{abstract}

Our understanding of magnetic reconnection (MR)  under chromospheric conditions remains limited. Recent observations have demonstrated the important role of ion-neutral interactions in the dynamics of the chromosphere. 
Furthermore, the comparison between spectral profiles and synthetic observations of reconnection events suggest that current MHD approaches appear to be inconsistent with observations. First, collisions and multi-thermal aspects of the plasma play a role in these regions. Second, hydrogen and helium ionization effects are relevant to the energy balance of the chromosphere.

This work investigates multi-fluid multi-species (MFMS) effects on MR in conditions representative of the upper chromosphere using the multi-fluid Ebysus code. We compare an MFMS approach based on a helium-hydrogen mixture with a two-fluid MHD model based on hydrogen only. The simulations of MRs are performed in a Lundquist number regime high enough to develop plasmoids and instabilities. We study the evolution of the MR and compare the two approaches including the structure of the current sheet and plasmoids, the decoupling of the particles, the evolution of the heating mechanisms, and the composition. 

The presence of helium species leads to more efficient heating mechanisms than the two-fluid case. This scenario, which is out of reach of the two-fluid or single-fluid models, can reach transition region temperatures starting from upper chromospheric thermodynamic conditions, representative of a quiet Sun scenario. The different dynamics between helium and hydrogen species could lead to chemical fractionation and, under certain conditions, enrichment of helium in the strongest outflows. This could be of significance for recent observations of helium enrichment in the solar wind in switchbacks and CMEs.

\end{abstract}


\section{Introduction}

Several plasma state transitions are found in the chromosphere. Indeed, the plasma can be partially or fully ionized, weakly or strongly magnetized, weakly or strongly collisional, all depending on the region considered \citep[see][]{Vernazza:1978uq,Martinez_Sykora_2020,Ni_2020}. Furthermore, the solar atmosphere is a complex environment composed of many species with a large range of ionization levels. In partially ionized plasmas, charged species interact with other charged and neutral particles through collisions. Recent studies have suggested that ion-neutral interaction effects play an important role in the dynamics of the low solar atmosphere and its energy balance \citep[see][and references therein]{Martinez-Sykora:2015fv,Ballester:2018fj,Soler2022}. Recent observations show that many dynamic and heating events, including through episodic magnetic reconnection, occur at chromospheric and transition region heights \citet{De-Pontieu:2009fk,De-Pontieu:2014fv, Peter:2014fk,Hansteen:2017rs,Guo_2020}.

Magnetic reconnection (MR) is a relaxation of the magnetic field's topology created by a local change in the magnetic field connectivity as described by \citet{Parker:1955rc}. This process is due to relevant dissipative effects in localized regions, called current sheets, where the magnetic field lines are reconnecting. This phenomenon is continuously present in the solar chromosphere \citep[][]{Gontikakis2013,Bhartu2013,Cargill2013,Klimchuk2015} as revealed by a wide range of observations, including the Interface Region Imaging Spectrograph \citep[IRIS,][]{De-Pontieu:2014yu} and the Swedish 1-m Solar Telescope (SST) telescopes \citep[e.g.,][]{Peter:2014fk,Guo_2020,Rouppe-van-der-Voort:2017uo} which provide unique observations of the dynamic outer solar atmosphere. In particular, recent observations with IRIS and SST \citep{Rouppe-van-der-Voort:2017uo,Young:2018bc} of strong brightenings in Si~{\sc iv} emission lines, ``UV bursts'', reveal evidence for the formation of plasmoids in the presence of MR events. Observations with SST reveal the presence of blob-like features at Alfv\'enic speed in the Ca~{\sc ii} K line.
However, the comparison between spectral profiles of observed plasmoids and synthetic observations of reconnection suggests that 1- the number of plasmoids is large and 2- the observed motions within the reconnection site are much more complicated than those predicted by a classical single-fluid MHD approach, \citep[][]{Nobrega-Siverio:2017sim}. UV bursts may span over extended regions from the upper photosphere, the chromosphere to the transition region (TR), thus, collisional and multi-thermal aspects of the plasma may play a role and need to be investigated \citep[][]{Hansteen:2019AA...626A..33H}.

On the one hand, several studies try to approximate collisions and multi-fluid aspects of the plasma by including the generalized Ohm's Law with the ambipolar diffusion, see for instance \citet{Leake2006,Khomenko:2018rm,Martinez_Sykora_2020,Rempel_2021}. This approach is generally used to simulate the solar atmosphere and accounts for ion-neutral interactions while retaining the simplicity of a single-fluid approach. On the other hand, several two-fluid MHD models \citep[e.g.,][]{Alvarez-Laguna:2016ty,Ni_2018,Braileanu2019,Wojcik2020,Ni_2020,Pelekhata2021,Niedziela2021,Leake:2013dq} have been constructed to investigate collisions and multi-thermal aspects of the plasma in the solar chromosphere. In that framework, the two fluids can interact through collisions and reactions, while the ions and electrons can interact with the electromagnetic field. Alternatively, the multicomponent model \citep[e.g.,][]{graille,Wargnier_2018,wargnierthesis2019} has been derived from kinetic theory and developed from a multi-scale analysis of the non-dimensional Boltzmann equations for electrons and heavy species using a generalized Chapman-Enskog expansion. Note that we refer to heavy species or fluids to any species except electrons. This model is a sound alternative to two-fluid models and is suitable for both partially- and fully-ionized multicomponent plasmas. Moreover, it accounts for the thermal non-equilibrium between the heavy particles (any ions and neutrals) and the electrons with a rigorous and detailed description of the dissipative effects based on kinetic theory of gases.

In \cite{Leake:2012pr,Leake:2013dq}, a study of MR with a  two-fluid simulation of MR for a weakly ionized reacting plasma, focusing on the solar chromosphere by considering collisions, thermal non-equilibrium between ions and neutrals, ionization/recombination, as well as optically thin radiative losses, was performed. The studies cited above focus on a two-fluid approach based on a single species with one ionized level of hydrogen. However, the Sun has additional species (such as helium and ionized metals) all with several excited and ionized levels with different degree of magnetization and collisional cross-sections that potentially could change the thermodynamics of the MR and its topology. Additionally, recent studies show evidence of the importance of helium ionization in the energy balance of the solar chromosphere, as shown in \citet{Hansteen:1997ag,Golding_2014,Golding_2016,Golding_2017,Martinez_Sykora_2020}. It has been demonstrated that the helium ionization state is often far from Local Thermodynamic Equilibrium (LTE) and that models require that both the temporal state and photo- and collisional-ionization effects are considered in order to be consistent with observations. Therefore, the most recent multi-fluid simulations of reconnection lack the self-consistent treatment of the relevant physical processes involved in the solar atmosphere, such as decoupling between various types of species, like hydrogen or helium species, including their respective ionized and excited levels.

In solar chromospheric conditions, and also in other astrophysical environments \citep[][]{gosling2007observations,pucci_velli_2020}, MR processes mostly evolve in a high Lundquist number regimes. In this regime, current sheets become rapidly unstable to resistive tearing instabilities which produces plasmoids \citep[e.g.,][]{Bhattacharjee2009,bhattacharjee2010,pucci_velli_2020}. Plasmoids are believed to play a major role in speeding up reconnection by having an influence on the variation of the current sheet size, as shown by \citet{Murtas2021}. Under the formation of plasmoids, the current sheet breaks into fragments or sections \citep[see][]{Leake_2020} and the resulting high current densities in each of these sections facilitate a high reconnection rate, as shown by \citet{Leake:2012pr,Leake:2013dq}. Plasmoid formation due to the instability of Sweet-Parker current sheets has been extensively examined through numerical studies with single-fluids  \citep[][]{Guo_2020,Loureiro_2015,Stenolfson1984,Ugai1995,Loureiro_2005,Leake_2020}. Recent works have proved that in fully ionized plasmas it is possible to trigger plasmoids for a critical Lunquist number $S_c$ ranging from $10^3$ to $10^4$, as described by \citet{Bhattacharjee2009,Cassak2009,Huang2010,Ni2010}. However, many studies have also demonstrated that this critical Lundquist number may vary by several orders of magnitude depending on the initial setup, such as the current sheet configuration, the amplitudes of viscosity and perturbations, and the plasma beta coefficient $\beta_p$, \citep[see][]{Ni2012,Ni2013,Huang_2017}. Additionally, the role of partial ionization on the onset of MR and development of the resistive tearing instability has been demonstrated \citet{zweibel2011magnetic,Pucci_2020,Murtas2021}. However, most of these studies consider only hydrogen species and do not consider additional species such as helium. Therefore, we aim to expand the studies of plasmoid formation and instabilities by including several species.

Recent measurements from SWEAP and FIELD on board of Parker Solar Probe \citep{Kasper_2016} revealed that Alfvénic magnetic field reversals called switchbacks (SBs) are associated with a significant increase in alpha particle abundances at the upper layers of the solar atmosphere \citet{Bale_2021}. Even though the SBs origin is not clear and could be associated with different possible mechanisms (magnetic reconnections, nanoflares, wave dissipation, footpoint shearing, or turbulence), observed SBs are characterized by strong periodic spatial modulations of the amplitude of the magnetic fields combined with large increase of alpha particles at coronal height. This spatial characteristic scale is consistent with solar super-granulation scales suggesting that the driver is connected to the super-granular network magnetic field \citep{Bale_2021,Fargette_2021}. These are also distinct features of slow solar wind \citet{woolley2021}. The chemical fractionation involving helium species should occur in the chromosphere, where plasma is partially ionized, suggesting the driver mechanism of these phenomena may be connected to this region. Addressing the chemical fractionation could help us understand the switchbacks' drivers and slow solar wind, and may provide insight into the enhancement of alpha particles in the upper layers of the solar atmosphere. 

In this work, a magnetic reconnection event in conditions typical of the upper chromosphere is studied. To do so, we consider a Lundquist number environment greatly above the critical Lundquist number $S_c\approx4\times10^3$ that leads to an instability regime characterized by the production of plasmoids, see \citet{Murtas2021} and references therein. The multi-fluid multi-species (MFMS) model accounts for multiple species with different ionized and excited levels, as described by \cite{Martinez_Sykora_2020}. We aim to study many fundamental physical processes and assess their impact on the reconnection process. By focusing on one scenario of a magnetic reconnection event, we perform two simulations based on two different mixtures: a two-fluid (TF) hydrogen mixture based on H, H$^+$ and electrons and a multi-fluid multi-species (MFMS) hydrogen helium mixture based on H, H$^+$, He, He$^+$ and electrons. This allows highlighting the impact of helium species with different ionized levels on the reconnection process. Additionally, we assess the impact of the MFMS and two-fluid approach on the structure of plasmoids and their formation.

The structure of this paper is as follows. In \refsec{sec:model}, the MFMS and TF model and their assumptions are described. \refsec{sec:numerical} briefly describes the numerical methods considered in \ebysus.  The initial conditions of the MRs simulations are inspired by upper-chromospheric conditions (\refsec{sec:methodology}). In \refsec{sec:2drec}, we analyze the results of the numerical simulations based on MFMS and TF models at different levels: the structure of the current sheet and plasmoids, the velocity fields and decoupling between species, the evolution of the heating processes and composition of the plasma. Finally, Section~\ref{sec:con} contains the discussion and the summary.

\section{Multi-fluid multi-species and two-fluid governing equations}~\label{sec:model}
We briefly summarize the two-fluid (TF) and multi-fluid multi-species (MFMS) MHD governing equations and their assumptions. Note that all equations are written in SI units. For MFMS, we focus on a hydrogen-helium mixture $\mixture$ such as $\mixture=\heavy\cup\{\elec\}$, where $\heavy=\left\{\text{H},\text{H}^+, \text{He}, \text{He}^+\right\}$ and $\elec$ denotes the set of heavy particles and electrons, respectively. Whereas TF has a pure hydrogen mixture $\mixture_{\text{TF}}=\heavy_{\text{TF}}\cup\{\elec\}$ where $\heavy_{\text{TF}}=\left\{\text{H},\text{H}^+\right\}$. No excited levels will be considered in this work.  $\mathfrak{n}_{\mathrm{\mixture}}$ and $\mathfrak{n}_{\mathrm{\heavy}}$ denote the number of species in  $\mixture$ and $\heavy$, respectively.

 For clarity and consistency, we use a similar nomenclature used by \citet{Ballester:2018fj,Khomenko:2014nr} with minor adjustments. The ionization states are referred as $\ions$, \ie\ $\ions=0$ denotes neutrals and $\ionized = \ions \geq 1$ ions. The identity of the chemical species (here hydrogen or helium) is indicated by $\specie$. Consequently, each set of particles in a given micro-state is described by $\specie\ions\in\mixture$ or $\specie\ions\in\mixture_{\text{TF}}$.  For electrons, the notation $\specie\ions$ is reduced to just $\{\elec\}$. In both TF and MFMS models, we neglect the electron inertia and assume quasineutrality. Additionally, we neglect the viscous effects associated with electrons. Each species has its own temperature $T_{\specie\ions}$ and no internal degrees of freedom. Finally, we neglect the effect of gravity.
 
 In the following, the governing equations will be written for the general case, \ie\, for MFMS based on $\mixture$. The set of equations for TF based on $\mixture_{\text{TF}}$ will be presented as a simplification of MFMS.

\subsection{Continuity Equations for Heavy Particles}~\label{sec:cont}

The mass density for each type of fluid in a given micro-state $\specie\ions\in\heavy$ is governed by the continuity equation in this generic form: 

\begin{eqnarray}
\dt \rho_\mathrm{\specie\ions}+ \nabla \cdot \left(\rho_\mathrm{\specie\ions}\,  {\velocity_\mathrm{\specie\ions}}\right) =&\nonumber \\ m_\mathrm{\specie\ions}  &(\Gamma^{\text{ion}}_{\mathrm{\specie\ions}}+\Gamma^{\text{rec}}_{\mathrm{\specie\ions}}),\quad \specie\ions\in\heavy, 
\label{eq:continuity} 
\end{eqnarray}

\noindent where $\rho_\mathrm{\specie\ions}=m_\mathrm{\specie\ions}\, n_\mathrm{\specie\ions}$ and $m_\mathrm{\specie\ions}$ are the mass density and particle mass respectively, $\velocity_\mathrm{\specie\ions}$ is the velocity, $\Gamma^{\text{rec}}_{\mathrm{\specie\ions}}$, and $\Gamma^{\text{ion}}_{\mathrm{\specie\ions}}$ are the mass transition rate due to recombination or de-excitation, and ionization or excitation, respectively. For any $\specie\ions\in\heavy$, these rates are defined as 

\begin{equation}
\Gamma^{\text{ion}}_{\mathrm{\specie\ions}} =\sum_{\specie\ions'\in\heavy\neq\specie\ions}\begin{cases}
      n_\mathrm{\specie\ions'}  \nu_{\mathrm{\specie \ions,\ions'}}^{\text{ion}}, & \text{if}\, \ions>\ions'\\
      -n_\mathrm{\specie\ions}  \nu_{\mathrm{\specie \ions,\ions'}}^{\text{ion}}, & \text{otherwise,}
    \end{cases} 
\end{equation}
and 
\begin{equation}
\Gamma^{\text{rec}}_{\mathrm{\specie\ions}} = \sum_{\specie\ions'\in\heavy\neq\specie\ions}\begin{cases}
      -n_\mathrm{\specie\ions}  \nu_{\mathrm{\specie \ions,\ions'}}^{\text{rec}}, & \text{if}\, \ions>\ions'\\
      n_\mathrm{\specie\ions'}  \nu_{\mathrm{\specie \ions,\ions'}}^{\text{rec}}, & \text{otherwise,}
    \end{cases}
\end{equation}

\noindent where $\nu_{\mathrm{\specie \ions,\ions'}}^{\text{ion}}$ and $\nu_{\mathrm{\specie \ions,\ions'}}^{\text{rec}}$ are the ionization and recombination frequencies between particles with micro-state $\mathrm{\specie\ions}$ and $\mathrm{\specie\ions'}\in\heavy$. In both MFMS and TF, the frequencies associated with the transition rates $\nu^{rec}$ and $\nu^{ion}$ include radiative and dielectronic recombination where the rates have been taken from \cite{Aldrovandi}. The ionization and three-body recombination rates have been taken from \cite{Janev_1989}. Photoionization is roughly modeled by simply assuming constant rates taken from \cite{Vernazza:1981yq} for the H-H$^+$ and He-He$^+$ transitions. Note that for TF the continuity equations presented in Eq.~\ref{eq:continuity} are identical to MFMS by simply changing the set of heavy particles $\heavy$ by $\heavy_{\text{TF}}$.

\subsection{Momentum Equations for Heavy Particles}~\label{sec:mom}

In MFMS, the momentum equation associated with a given species defined by its microstate $\specie\ions\in\heavy$  is defined as follows:

\begin{eqnarray}
\dt (\rho_\mathrm{\specie\ions} {\velocity_\mathrm{\specie\ions}}) &&+ \nabla \cdot \left(\rho_\mathrm{\specie\ions} {\velocity_\mathrm{\specie\ions}}\otimes {\velocity_\mathrm{\specie\ions}} + P_\mathrm{\specie\ions} \identity-\viscous_{\specie\ions}\right)  =  \nonumber \\
&&   n_\mathrm{\specie\ions} q_\mathrm{\specie\ions} \left(\EField + {\velocity_\mathrm{\specie\ions}} \pvec \BField \right)  \nonumber \\
&& +\momex^{\text{ion}}_\mathrm{\specie\ions} +\momex^{\text{rec}}_\mathrm{\specie\ions} +\momex_\mathrm{\specie\ions}^{\text{col}},\quad \specie\ions\in\heavy \label{eq:spmoma1}
\end{eqnarray}

\noindent where $q_{\specie\ions}$ and $P_{\specie\ions}$ are the ion charge and gas pressure of a given species $\specie\ions\in\heavy$, $\viscous_{\specie\ions}$ is the viscous stress tensor associated with species $\specie\ions\in\heavy$, $\EField$ and $\BField$ are the electric and magnetic field, $\momex^{\text{ion}}_\mathrm{\specie\ions}$ and  $\momex^{\text{rec}}_\mathrm{\specie\ions}$ are the changes in momentum for species $\specie\ions$ due to ionization and recombination processes respectively, $\momex_{\specie\ions}^{\text{col}}$ is the sum of all the momentum exchange between particles with a given micro state $\specie\ions\in\heavy$ with any other particles with another micro state $\specie'\ions'\in\mixture$. 

In TF, the momentum equation written in Eq.~\ref{eq:spmoma1} slightly differs. We solve for 1- a momentum equation for neutrals (here H) and 2- a momentum equation for all ionized species that is the sum of the momentum equations for ions (here $\text{H}^+$) and electrons. Therefore, the Lorentz force, corresponding to the first right-hand side term of Eq.~\ref{eq:spmoma1}, reads $\totalcurrent \pvec \BField $ where $\totalcurrent$ is the total current density. Additionally, in the momentum equations for ions, the pressure term in the convective fluxes is simply replaced by the total pressure of charged particles $\nabla\cdot\left(\left[P_{\text{H}^+}+P_{\elec}\right]\identity\right)$. Note that in MFMS the term associated with the gradient of pressure of the electrons is involved in the definition of $\EField$.

In this study, we consider plasmas evolving in a high Reynolds number regime leading to characteristic length scales associated with viscous effects that are generally much smaller than any chosen grid size. In this context, an artificial subgrid-scale model of these viscous terms $\viscous_{\specie\ions},\,\specie\ions\in\mixture$ or $\mixture_{\text{TF}}$ has been considered in order to represent shocks and turbulence. This approach is inspired from \citet{Nordlund1982,Gudiksen:2011qy} and has been extended for MFMS and TF models. 

In MFMS, the sum of all the momentum exchange for a fluid of a specific particle with a given micro state $\specie\ions\in\heavy$ is defined as: 

\begin{eqnarray}
\momex_{\specie\ions}^{\text{col}}&&=\sum_{\specie'\ions'\in\mixture}\momex^{\text{col}}_{\specie\ions,\specie'\ions'}  \, \nonumber\\&&= \sum_{\specie'\ions'\in\mixture}m_{\specie\ions}\, n_{\specie\ions}\nu^{\text{col}}_{\specie\ions,\specie'\ions'} ( \velocity_{\specie'\ions'} - \velocity_{\specie\ions}),\; 
\label{eq:momexhp}
\end{eqnarray}
 where $\nu^{\text{col}}_{\specie\ions,\specie'\ions'}$ is the collision frequency between particle $\specie\ions$ and $\specie'\ions'$. The momentum exchange operators are symmetric, \ie\,  $\forall\,(\specie\ions, \specie'\ions')\in\heavy\times\mixture,\;\momex^{\text{col}}_{\specie\ions, \specie'\ions'} = - \momex^{\text{col}}_{\specie'\ions',\specie\ions}$. Note that Eq.~\ref{eq:momexhp} does not include the term associated with the difference in heat fluxes between species, as described by \citet{Hansteen:1997ag} or in the 13N moment model of \citet{zhdanov_2002}. The collision frequencies $\nu^{\text{col}}_{\specie\ions, \specie'\ions'}$ are described in detail in \citet{Wargnier2022}. In this context, the collisional frequencies are consistent with the definition given by \citet{zhdanov_2002} in the context of the 13-N moment model for multicomponent plasmas, calculated from generalized Chapman-Cowling collision integrals. The collision integrals data for $\mixture$ and $\mixture_{\text{TF}}$ have been taken from \citet{brunoetal2010} based on a review of experimental measurements. These collision integrals include charge-exchange interactions for H-H$^+$ and He-He$^+$ interactions.

\subsection{Thermal Energy Equations}~\label{sec:ener}

In both MFMS and TF models, we account for thermal decoupling between all particles, including electrons. Therefore, we solve for the equations of thermal energy for each particle $\specie\ions\in\mixture$ or $\mixture_{\text{TF}}$.

In MFMS, the energy equations for each heavy particle and electrons read as follows,

\begin{eqnarray}
 \dt e_\mathrm{\specie\ions}  +\nabla\cdot&&\left( e_\mathrm{\specie\ions}\velocity_\mathrm{\specie\ions}\right)+P_\mathrm{\specie\ions}\nabla\cdot\velocity_\mathrm{\specie\ions}= \nonumber \\
&&  Q^{\text{ion}}_\mathrm{\specie\ions} + Q^{\text{rec}}_\mathrm{\specie\ions} +   Q_\mathrm{\specie\ions}^{\text{col}} + Q_\mathrm{\specie\ions}^{\viscous},\quad \specie\ions\in\heavy \label{eq:enei5}
\end{eqnarray}

\begin{eqnarray}
\dt e_\mathrm{\elec} +\nabla\cdot&& \left(e_{\elec}\velocity_{\elec}\right) +P_\elec\nabla\cdot\velocity_\elec= \nonumber \\ 
&&Q^{\text{ion}}_\elec + Q^{\text{rec}}_\elec +Q_\mathrm{\elec}^{\text{col}}+Q_\mathrm{\elec}^\mathrm{spitz},\label{eq:enee5} 
\end{eqnarray}

\noindent where $e_{\specie\ions}$ and ${e_\elec}$ are the thermal energies of any heavy particles $\specie\ions\in\heavy$ and electrons respectively. $Q^{\text{ion}}_{\specie\ions}+Q^{\text{rec}}_{\specie\ions}$ and $Q^{\text{ion}}_{\elec}+Q^{\text{rec}}_{\elec}$ are the heating/cooling term due to the ionization and recombination processes associated with particles $\specie\ions\in\heavy$ and electrons respectively, $Q_{\elec}^\mathrm{spitz}$ is the heating term due to the electron thermal conduction along the magnetic field, $Q_\mathrm{\specie\ions}^{\viscous}$ is the heating term due to viscous effects associated with particle $\specie\ions\in\heavy$, $Q_{\specie\ions}^{\text{col}}$ and $Q_{\elec}^{\text{col}}$ are the heating terms due to collisions associated with heavy particle $\specie\ions\in\heavy$ and electrons respectively. In TF, the energy equations Eqs.~\ref{eq:enei5} and \ref{eq:enee5} are obtained by substituting $\specie\ions\in\heavy_{\text{TF}}$ to $\specie\ions\in\heavy$. 

Note that  $Q^{\text{ion}}_\mathrm{\elec}$ corresponds to the optically thin radiative losses term which involve hydrogen and helium ionization potential coefficients, as described by \citet{Leake:2012pr,Leake:2013dq}. The Spitzer term is anisotropic with respect to the magnetic field and is identical to the definition from \citet{Spitzer56,Gudiksen:2011qy}. The heating term due to viscous effects $Q_\mathrm{\specie\ions}^{\viscous}$ is derived from the viscous terms in momentum equations $\viscous_{\specie\ions}$ leading to a similar expression introduced by \citet{Nordlund1982,Gudiksen:2011qy} for a single-fluid MHD model.

The collisional terms associated with any particle $\specie \ions\in\mixture$ are defined  as 

\begin{equation}Q_{\specie\ions}^{\text{col}} = \sum_{\specie'\ions'\in\mixture} Q_{\specie\ions,\specie'\ions'}^{\text{col}} =\sum_{\specie'\ions'\in\mixture} Q_{\specie\ions,\specie'\ions'}^{\textbf{u},\text{col}}+Q_{\specie\ions,\specie'\ions'}^{T,\text{col}}.
\label{eq:collisionheat}
\end{equation}

The first component of Eq.~\ref{eq:collisionheat} corresponds to the heating produced by the drift velocity between species $(\specie\ions,\specie'\ions')\in\mixture^2$ due to collisions. The second component corresponds to a thermalization process where temperature $T_{\specie\ions}$ relaxes towards $T_{\specie'\ions'}$ at a collisional rate $\nu^{\text{col}}_{\specie\ions,\specie'\ions'}$. The heating produced by the drift velocity and thermalization terms between species $\specie\ions$ and $\specie'\ions'$ are defined as
\begin{equation}
\begin{split}
Q_{\specie\ions,\specie'\ions'}^{\textbf{u},\text{col}} &= \nu^{\text{col}}_{\specie\ions, \specie'\ions'}\,m_{\specie\ions,\specie'\ions'}n_{\specie\ions}| \velocity_{\specie'\ions'} - \velocity_{\specie\ions}|^2,\\
Q_{\specie\ions,\specie'\ions'}^{T,\text{col}}&= \nu^{\text{col}}_{\specie\ions \specie'\ions'}\left(\tfrac{3 m_{\specie\ions}}{m_{\specie\ions}+m_{\specie'\ions'}}\right)n_{\specie\ions}k_B\left(T_{\specie'\ions'}-T_{\specie\ions}\right).
\end{split}
\label{eq:heatingvelocitydrift}
\end{equation}

Note that for any $(\specie\ions,\specie'\ions')\in\mixture^2$ or $\mixture^2_{\text{TF}}$, thermalization terms $Q_{\specie\ions,\specie'\ions'}^{T,\text{col}}$ do not contribute to the total thermal energy (sum of all the thermal energy equations Eqs.~\ref{eq:enei5} and \ref{eq:enee5}) of the multi-fluid plasma since these terms are symmetric with respect to species. Similarly, for any $\specie\ions\in\heavy$ or $\heavy_{\text{TF}}$, ionization/recombination terms of heavy species $Q^{\text{ion}}_\mathrm{\specie\ions} + Q^{\text{rec}}_\mathrm{\specie\ions}$ do not contribute to the total thermal energy. However, the heating produced by drift velocities, optically thin radiative losses and Spitzer terms contribute to the total thermal energy.

 

\subsubsection{Equation of States}

 In both MFMS and TF, we assume equation of states for electrons and heavy species following classical ideal gases laws such as  

\begin{equation}
P_\mathrm{\specie\ions}=n_\mathrm{\specie\ions}k_B T_\mathrm{\specie\ions} = (\gamma-1) e_\mathrm{\specie\ions},\quad\specie\ions\in\heavy \,\,\,\text{or}\,\,\, \heavy_{\text{TF}}
\end{equation}
and 

\begin{equation}
P_{\mathrm{\elec}}=n_{\mathrm{\elec}} k_B T_{\mathrm{\elec}} = (\gamma-1) e_{\mathrm{\elec}}, 
\end{equation}

\noindent where $\gamma=5/3$ is the adiabatic constant taken as identical for all species.

\subsection{Generalized Ohm's law and magnetic Induction Equation}~\label{sec:elect}

In both MFMS and TF, we compute the electric field from a simplified electron momentum equation in the asymptotic limit  where we assume that 1- the electron inertia and its time variation is neglected, 2- the mass of electrons is much smaller than the mass of any heavy particles and 3- the effects associated with ionization and recombination are neglected. In this framework, in both MFMS and TF, the Ohm's law reads

\begin{eqnarray}
\EField = - \velocity_{\elec}\pvec\BField + \frac{1}{n_\elec q_\elec}\left[\nabla P_{\elec}
-\momex_{\elec}^{\text{col}}\right],  \label{eq:e_ohm} 
\end{eqnarray}

\noindent where $\momex_{\elec}^{\text{col}}$ has the same definition as Eq.~\ref{eq:momexhp} by replacing a given heavy species $\specie\ions\in\heavy$ by $\specie\ions=\{\elec\}$. The electromagnetic field is not considered as an external force but their evolution is governed by the set of Maxwell equations. Therefore, combining Eq.~\ref{eq:e_ohm} with Maxwell-Faraday equation, 
\begin{equation}
  \dt\BField = - \nabla\pvec\EField, \label{eq:magne}  
\end{equation}
we obtain the magnetic induction equation. 

In MFMS, since several ionized species are considered and the quasi-neutrality approximation is assumed, the electron velocity can be expressed as a function of the hydrodynamic velocity of each ion and the total current as follows

\begin{equation}
    \velocity_{\elec} = \sum_{\specie\ions}\frac{n_{\specie\ions}q_{\specie\ions}\velocity_{\specie\ions}}{n_{\elec}q_{\elec}}-\frac{\totalcurrent}{n_{\elec}q_{\elec}}.
    \label{eq:vele}
\end{equation}

Note that the contribution of the total current $\totalcurrent$ of the electron velocity in the equation of the electric field in Eq.~\ref{eq:e_ohm}, also known as the Hall term, has not been considered for this study (however, we will investigate the effect of this term in future studies). In this study, since the characteristic speed of each particle is much smaller than the speed of light in vacuum, the Maxwell-Ampere's law can be simplified by neglecting the displacement current term, which leads to the following relation of the total current
\begin{equation}
\totalcurrent = (\nabla \pvec \BField)/\mu_0,
\end{equation}
where $\mu_0$ is the vacuum permeability. Note that from Eq.~\ref{eq:e_ohm} it is possible to identify one component of the resistivity of the MFMS model as function of collision integrals, see Section~\ref{sec:resistivity} for further details.

The set of equations associated with MFMS and TF described in Eqs.~\ref{eq:continuity},\,\ref{eq:spmoma1},\,\ref{eq:enei5},\,\ref{eq:enee5} and \ref{eq:magne} is a closed system of equations which can be solved and integrated with a specific numerical strategy.


\section{Numerical methods}~\label{sec:numerical}
 
The two sets of governing equations for MFMS and TF models presented in Eqs.~\ref{eq:continuity}, \ref{eq:spmoma1}, \ref{eq:enei5}, \ref{eq:enee5}, and \ref{eq:magne} have been implemented in a code named \ebysus\, \citep{Martinez_Sykora_2020}, which is an extension of the single-fluid radiative MHD code \bifrost\, \citep{Gudiksen:2011qy}. Note that the governing equations can be rewritten in a more compact form which allows us to distinguish the nature of the different terms involved, as shown in Appendix~\ref{app:system}. 

Concerning the spatial discretization, a sixth-order differential operator has been considered for the convective fluxes, as in \bifrost. The conservative quantities are not co-located in space and are considered on a staggered grid, thus, a fifth-order interpolation is used to relocate variables as needed. Since a high-order finite difference scheme has been considered, it is convenient to numerically stabilize shocks and discontinuities with artificial numerical terms. In the literature, several type of artificial terms have been introduced, as described by \citet{WILKINS1980281,COOK2005379} and references therein. By inheritance of the \bifrost\, code, the structure of the artificial terms are the hyperdiffusive terms described by \citet{Nordlund1982,Gudiksen:2011qy}. These terms are considered in order to capture shocks and discontinuities while treating correctly turbulence in a high Reynolds number regime. In \ebysus\, the hyperdiffusive terms have been expanded for the multi-fluid case from those considered in \bifrost\, since they depend on the velocity and speeds of sound for each species and the Alfven speed of all ions. The constant coefficients introduced in these terms have been chosen based on fits to a wide variety of shock and turbulence test problems.

The integration of the two presented systems requires some specific temporal integration methods. Indeed, these systems can be seen as convective-diffusive systems with stiff source terms \citep[see][]{ABDULLE2013869,duarte2013,descombes2004}, as presented in Appendix~\ref{app:system}. Some of these source terms (in particular those associated with collisions and ionization/recombination) are numerically stiff since they are associated with physical processes where timescales often are much smaller than the convective timescales. Therefore, these terms are subject to strong CFL constraints leading to small timesteps and drastically increased computational costs. To avoid this, our strategy is to consider a first order Lie splitting approach in order to split the temporal integration on different subsystems. First, the convective terms have been integrated following a third-order predictor-corrector Hyman \citep{hyman1979}. Alternately, it is possible to consider a third-order explicit Runge-Kutta method. Then, the diffusive term (\ie\, the Spitzer term) has been integrated by solving a non-Fickian (hyperbolic) diffusion equation as described by \citep[see][]{Rempel_2016}. Note that, in \ebysus , it is possible to integrate this term following an implicit multi-grid method, as described in \citet{Gudiksen:2011qy}. Finally, the stiff source terms (see $S^{\text{stiff}}$ in Appendix \ref{app:system}) are integrated following an Ordinary Differential Equation (ODE) solver based on a fifth-order implicit Runge Kutta method also known as \textit{Radau IIA} method \citep[see][]{HAIRER199993,Hairer1996}. In order to maintain the solenoidal constraint on the magnetic field, a Hodge projection method has been used \citep[see][]{Hodgemethod1980}.

\ebysus\, allows the introduction of as many species as desired using DIPER format atom files and the CHIANTI database \citep[see][]{Judge_2007,delzanna2015}. The number of ionized and excited levels, transitions, and rates are detailed in these atom files. The code allows one to easily perform numerical simulations with any chosen mixtures following identical numerical strategies focusing on the system of equations presented in the previous section, as long as the atom files associated with all species considered are provided. The \ebysus\, code also has the possibility of running a single-fluid MHD option allowing a comparison with multi-fluid MHD models using the same numerical code.
 
 
\section{Methodology and initial conditions}~\label{sec:methodology}

In this study, we focus on a magnetic reconnection configuration under specific thermodynamic conditions that aim to represent the upper solar chromosphere. For identical initial thermodynamic conditions, we perform a comparison between TF based on $\mixture_{\text{TF}}$ \citep[similar to what  was performed by][]{Alvarez-Laguna:2016ty,Leake:2012pr,Leake:2013dq} and MFMS based on $\mixture$. Our comparison highlights the impact of helium on the reconnection process.

\subsection{Initial reconnection setup}

The reconnection setup is inspired from \cite{Leake:2012pr,Leake:2013dq}, \ie\, a magnetic field configuration forming a Harris current sheet placed in a constant thermodynamic and plasma beta environment. In this work, we consider an initial temperature of $T_0=16\,000$ K and a total number density $n_{0}=7.47\times10^9$ $\text{cm}^{-3}$. Then, the population associated with each species involved in $\mixture$ and $\mixture_{\text{TF}}$ is calculated. In both MFMS and TF, the plasma beta coefficient is calculated from the total pressure of the plasma $P_0 = n_0 k_B T_0$ and is defined as $\beta_p=0.2$. We have considered a 2.5D domain based on a uniform grid, with $800 \times 800$ grid points where each grid cell size is$dy=dz=5.333$~km. In this configuration, the whole computational domain represents a 2.5D box of size $L_0=4.2664$ Mm. Open boundary conditions have been considered in all directions for all the variables and fields with linear extrapolations while minimizing strong gradients. In order to mitigate reflections, the last $0.1332$~Mm grid points inside the domain are dedicated for extrapolating the fields and damping waves that could eventually disturb the reconnection site.

The Harris current sheet, as described by \cite{Leake:2012pr,Leake:2013dq}, is defined as follows

\begin{equation} 
\mathbf{A_x} = -\frac{B_0}{2} \lambda_{\psi} \log\left(\cosh(\frac{y}{\lambda_{\psi}})\right)\mathbf{e_x},\; \mathbf{B} = \left[\nabla\wedge \mathbf{A_x}\right], 
\end{equation}

\noindent where $\lambda_{\psi}$ is the initial thickness of the current sheet defined as $\lambda_{\psi}=0.1 L_0$, $B_0$ is the amplitude of the magnetic field (in balance with the total pressure  $P_0 = \beta_p \frac{B_0^2}{2\mu_0}$) equal to $1.42$ G, $\mathbf{A_x}$ is the initial magnetic potential in the x direction, and $\mathbf{e_x}$ is the elementary vector in the x direction.

The pressure of the ionized species $P^{\text{ions}}$ is locally increased in the current sheet in order to balance the decrease of the Lorentz force in the current sheet, as follows: 
\begin{equation}
P^{\text{ions}}(y) = P^{\text{ions}}_{0} + \frac{ P_0^{\text{ions}}}{\cosh^2(\frac{y}{\lambda_{\psi}})},
\end{equation}
\noindent where $P^{\text{ions}}_{0}$ is the total pressure of the ions computed from the initial thermodynamic conditions ($T_0$ and $\rho_0$). 
In order to trigger the reconnection we have considered a local-centered perturbation on the magnetic field (while guaranteeing the divergence free of the magnetic field) as follows:

\begin{equation}
  \mathbf{B^{\alpha}_z} (y,z^*) = - \alpha(y-\frac{L_0}{2}) f(y,z^*)\mathbf{e_z}, \; (y,z^*)\in \left[0,L_0\right]^2
\end{equation}
\begin{equation}
    \mathbf{B^{\alpha}_y} (y,z^*) = \alpha (z^*-\frac{L_0}{2}) f(y,z^*)\mathbf{e_y}, \; (y,z^*)\in \left[0,L_0\right]^2
\end{equation}
\noindent where
\begin{equation}
    f(y,z^*) = \exp(-\left[\frac{z^*-L_0/2}{\lambda_{\psi}}\right]^2 )\exp(-\left[\frac{y-L_0/2}{\lambda_{\psi}}\right]^2 ),
\end{equation}
where $\alpha$ is the amplitude of the perturbation which has been chosen in such a way that the maximum amplitude of $|\mathbf{B^{\alpha}}|=\sqrt{|\mathbf{B^{\alpha}_y}|^2+|\mathbf{B^{\alpha}_z}|^2}$ corresponds to approximately $10\%$ of the amplitude of the magnetic field of the Harris current sheet $B_0$. This amplitude is chosen in accordance with \citet{Alvarez-Laguna:2016ty,Leake:2012pr}.

For both MFMS and TF, the characteristic initial Alfven speed is calculated as $V^{\text{Alf}}_0 = B_0/\sqrt{\mu_0  \rho^{\text{ions}}_0}=60$~km s$^{-1}$, where $\rho^{\text{ions}}_0$ is the total initial mass density of ions. The initial thermal speed is $V^{\text{therm}}_0=\sqrt{8 k_B T_0/\pi m_0}=18.3$~km s$^{-1}$, where $m_0$ is the characteristic mass chosen here as the mass of protons. No initial velocity field is considered, rather the velocity field is naturally triggered during the reconnection process. The numerical simulations based on MFMS and TF have been run until $t_{\text{end}}=454$~s and $325$~s, respectively. For the rest of this work, we will focus on a restricted domain in the $z$ direction where $z\in[0,4]$~Mm and considering the change of variable $z=z^*-\delta_z$ with $\delta_z=0.1332$~Mm.

\subsection{Initial composition}
\label{sec:composition}

\begin{table}[]%
\begin{center}
\begin{tabular}{c|cc|cc|c|c|}
\cline{2-7}
\multirow{2}{*}{}                               & \multicolumn{2}{c|}{H}        & \multicolumn{2}{c|}{H$^+$} & \multirow{2}{*}{He} & \multirow{2}{*}{He$^+$} \\ \cline{2-5}
                                                & \multicolumn{1}{l|}{$\mixture$} & $\mixture_{\text{TF}}$ & \multicolumn{1}{l|}{$\mixture$}       & $\mixture_{\text{TF}}$       &                     &                                        \\ \hline
\multicolumn{1}{|l|}{$n$ $\left[\%\right]$} &  \multicolumn{1}{c|}{34.55}  &    37.5  & \multicolumn{1}{c|}{57.55}        &     62.5       &    7.9                 &     $3.17\times10^{-6}$                                   \\ \hline
\multicolumn{1}{|c|}{$\rho$ $\left[\%\right]$}   & \multicolumn{1}{c|}{28}  &  37.5    &  \multicolumn{1}{c|}{46.6}        &    62.5        &         25.4             &                $10^{-5}$                         \\ \hline
\end{tabular}%
\caption{Initial number and mass densities in percentage for MFMS and TF. The initial number and mass densities percentage are calculated as $\alpha_{\specie\ions}\left[\%\right]=100\times\alpha_{\specie\ions}/\sum_{\specie\ions\in\beta}\alpha_{\specie\ions},\; \alpha\in \{n,\rho\}$ and $\beta\in\{\heavy,\,\heavy_{\text{TF}}\}$.}
\label{tab:table}
\end{center}
\end{table}

In both TF and MFMS, the initial number and mass densities for all species are calculated from the total number density $n_0$ and temperature $T_0$ assuming statistical and thermal equilibrium. The initial total number density of the plasma is the same for MFMS and TF but the total mass density differs due to the presence of neutral helium species in MFMS. The initial composition in number and mass densities for mixtures $\mixture$ and $\mixture_{\text{TF}}$ are given in Table~\ref{tab:table}. Note that, in MFMS, the initial abundance of helium species, \ie\ with respect to the total hydrogen number density, is approximately $8.5\%$, \ie\ photospheric abundances \citep{Asplund:2009uq}.

\subsection{Initial multi-fluid parameters}
\label{sec:mfluidparameters}

\begin{table*}[!htbp]
\centering
\begin{tabular}{l|l|l|l|l|l|l|l|l|}
\cline{2-9}
                                &  \multicolumn{1}{|c|}{H-H$^+$}&  \multicolumn{1}{|c|}{H-He}&  \multicolumn{1}{|c|}{H$^+$-He} &  \multicolumn{1}{|c|}{He$^+$-H} &  \multicolumn{1}{|c|}{He$^+$-H$^+$} &  \multicolumn{1}{|c|}{He$^+$-He} &  \multicolumn{1}{|c|}{$\elec$-H}&  \multicolumn{1}{|c|}{$\elec$-He} \\ \hline
\multicolumn{1}{|c|}{$\nu^{\text{col}}$~[s$^{-1}$]} & \multicolumn{1}{|c|}{$\approx 10\text{ -- }100$} &  \multicolumn{1}{|c|}{$\approx 0.1\text{ -- }1$} & \multicolumn{1}{|c|}{$\approx 0.1\text{ -- }1$} & \multicolumn{1}{|c|}{$\approx 0.1\text{ -- }1$} & \multicolumn{1}{|c|}{$\approx 100\text{ -- }1000$} & \multicolumn{1}{|c|}{$\approx 1\text{ -- }5$} &\multicolumn{1}{|c|}{$100\text{ -- }500$} & \multicolumn{1}{|c|}{$\approx 20\text{ -- }50$}  \\ \hline
\multicolumn{1}{|c|}{$\lambda_p$~[km]}       &\multicolumn{1}{|c|}{$\approx0.1\text{ -- }1$}  & \multicolumn{1}{|c|}{$\approx10\text{ -- }100$} & \multicolumn{1}{|c|}{$\approx10\text{ -- }100$} & \multicolumn{1}{|c|}{$\approx10\text{ -- }100$} & \multicolumn{1}{|c|}{$\approx0.01\text{ -- }0.1$} & \multicolumn{1}{|c|}{$\approx3\text{ -- }7$} & \multicolumn{1}{|c|}{$\approx2\text{ -- }3$}  & \multicolumn{1}{|c|}{$\approx15\text{ -- }30$}\\ \hline
\end{tabular}%
\caption{Initial range of values (minimum/maximum respectively) of collision frequencies in s$^{-1}$ and mean free path in km for MFMS in orders of magnitude. All the other interactions that are not shown in this table (for example $\elec$-H$^+$) have a mean free path largely below 0.01~km.}
\label{tab:table_colfreqHP}
\end{table*}

 In our initial setup, the rates associated with ionization or recombination processes have orders of magnitude ranging from $10^{-4}$ to $10^{-1.5}$~s$^{-1}$ for helium and hydrogen respectively.  For MFMS, the initial values of collisional frequencies and mean free paths (taken as the thermal speed of the colliding species times collisional frequencies between the two species) of the different interactions in orders of magnitude are presented in Table~\ref{tab:table_colfreqHP}. These values highlight the large disparities in the coupling between fluids due to collisions in MFMS. Note that in the TF model, the collisional rate between H and H$^+$ is slightly higher than in Table~\ref{tab:table_colfreqHP} for MFMS due to the initial mass density of H$^+$, which is higher in TF than in MFMS, as shown in Table~\ref{tab:table}.
 
 These preliminary calculations demonstrate that the different fluids involved in MFMS are expected to be strongly or weakly coupled due to collisions at the initial stage of the reconnection. For example, interactions involving He appears to be less coupled due to collisions than interaction H-H$^+$ by two orders of magnitude. However, interactions associated with He$^+$ and H$^+$ are more strongly coupled due to collisions than the H-H$^+$ interaction. Note that timescales associated with ionization and recombination are much slower than collisional timescales, thus, collisional processes are much faster than ionization and recombination processes.
 
One can notice that some of the characteristic mean free paths are not spatially resolved by the grid size $dy=dz$, in particular those associated with electrons. Therefore, the coupling due to collisions associated with electrons appears to be not fully resolved by the current numerical setup and would require a better spatial resolution. However, the chosen grid size numerically resolves the mean free paths associated with heavy particles within the simulation. Therefore, the following simulations resolve the coupling and decoupling of the various fluids except for electrons.

\section{Results}~\label{sec:2drec}

In this section, we analyze the results of our two simulations based on MFMS and TF models, keeping in mind that the mixture associated with MFMS is $\mixture=\left\{\text{H},\text{H}^+, \text{He}, \text{He}^+,\elec\right\}$, and the mixture associated with TF is $\mixture_{\text{TF}}=\left\{\text{H},\text{H}^+,\elec\right\}$. The analysis is performed at different levels. First, we focus on the structure of the current sheet and its evolution (Section \ref{sec:currentsheet}), and, then,  the velocity fields and highlight the decoupling between fluids (Section \ref{sec:decoupling}). We continue with Section \ref{sec:heating} in which we identify the main heating processes in both mixtures. Finally, we analyze the evolution of the ionization level in the current sheet in MFMS and TF and the chemical fractionation between helium and hydrogen species in MFMS during the reconnection process (Section \ref{sec:fractionation}).

\subsection{Current sheet and plasmoids}
\label{sec:currentsheet}

In both MFMS and TF, our calculations show that two phases of the reconnection process can be identified. Figure~\ref{fig:current} and associated animation illustrates the distribution of the orthogonal current $\totalcurrent_x$ and total velocity fields in both $y$ and $z$ directions for both MFMS and TF, at different times. The total velocity fields are calculated as $\velocity_{\text{tot}} = \left(\rho\velocity\right)_{\text{tot}}/\rho_{\text{tot}}$. Note that further details about each individual velocity are given in the next Sections~\ref{sec:decoupling}.

First, the simulation has a laminar or a Sweet-Parker reconnection phase, where only one elongated current sheet is formed and develops inflow and outflow velocity fields corresponding to the $y$ and $z$ direction, respectively \citep[see][]{Parker:1955rc,Parker:1963vn,Sweet1958} due to the initial structure of the magnetic field. This can be clearly identified for both MFMS and TF in the first column of Fig.~\ref{fig:current}, as well as in the animation corresponding to this figure. After this first phase, we have a plasmoid instability phase where many plasmoids are formed and are advected from the middle of the reconnection region towards the outflow region along the $z$ axis (the last column of Fig.~\ref{fig:current}). During this phase, the presence of plasmoids breaks the current sheet into multiple thinner current sheets, which reconnect faster than in the Sweet-Parker regime \citep[see][]{Murtas2021}. This is also in agreement with previous 2D \citep[\eg][]{Huang:2010PhPl...17f2104H} and 3D \citep[\eg][]{Huang:2016ApJ...818...20H} simulations.

At the very beginning of the laminar phase ($t\approx54$~s), the magnitude of the current inside the reconnection region is slightly higher in TF than in MFMS, as shown in the animation of Fig.~\ref{fig:current}. Additionally, the magnitude of the total velocity fields close to the current sheet and the magnitude of the total velocity in the $z$ and $y$ directions are slightly higher in TF than in MFMS. The current sheet is thinned sooner in TF than in MFMS, as shown in the animation of Fig.~\ref{fig:current} and in the top left panel of Fig.~\ref{fig:recorate}. At the end of the laminar phase and during the plasmoid instability phase, it remains difficult to compare the evolution of the current sheet and total velocity fields as it is unstable and fields change rapidly on short length scales. Therefore, further analysis has been performed and is presented in Fig.~\ref{fig:recorate}. 

\begin{figure*}[tbh]
\centering
    \hspace*{-1.7cm}
	\includegraphics[scale=0.4]{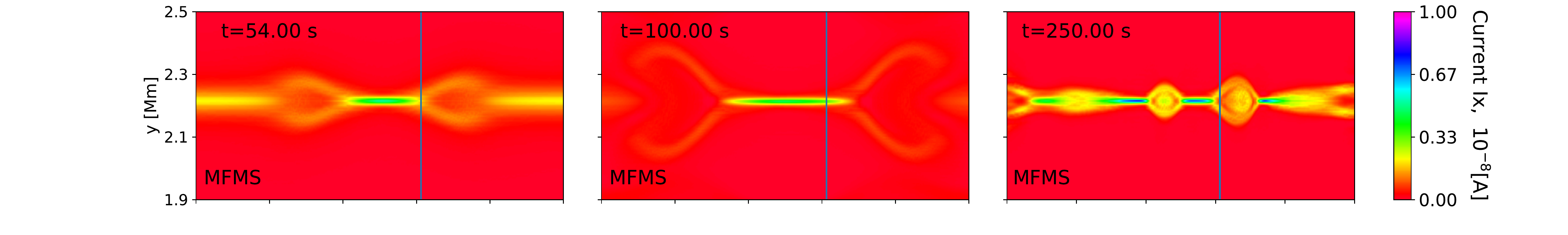}\vspace*{-0.2cm}\\
	\hspace*{-1.7cm}
	\includegraphics[scale=0.4]{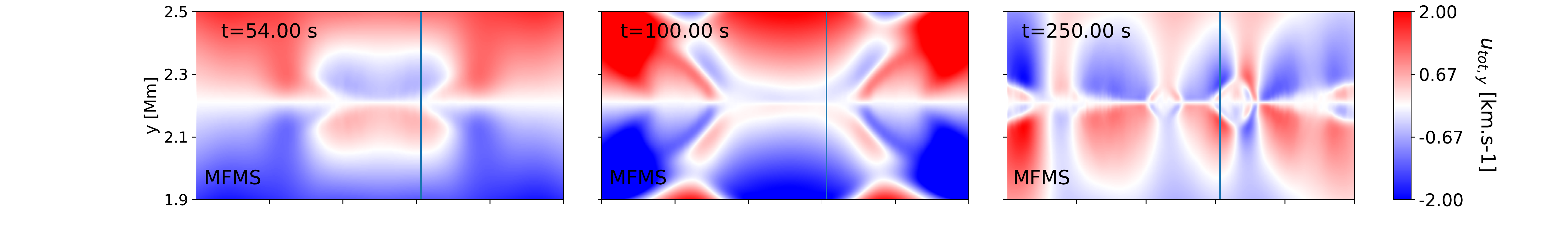}\vspace*{-0.2cm}\\
	\hspace*{-1.7cm}
	\includegraphics[scale=0.4]{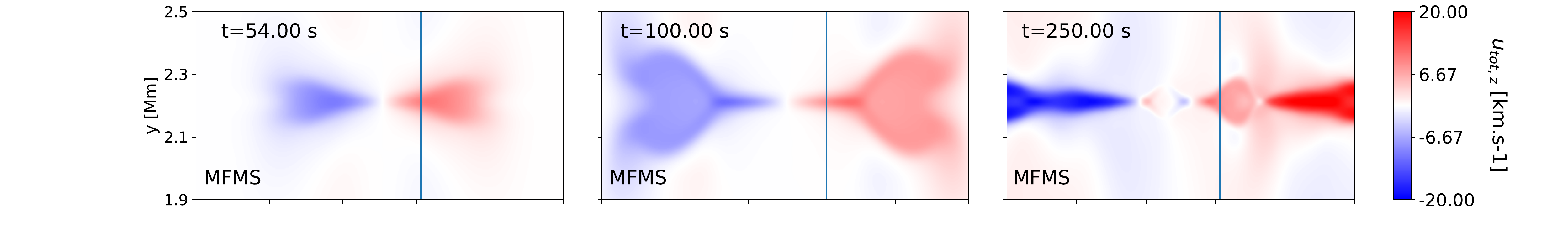}\vspace*{-0.2cm}\\
	\hspace*{-1.7cm}
	\includegraphics[scale=0.4]{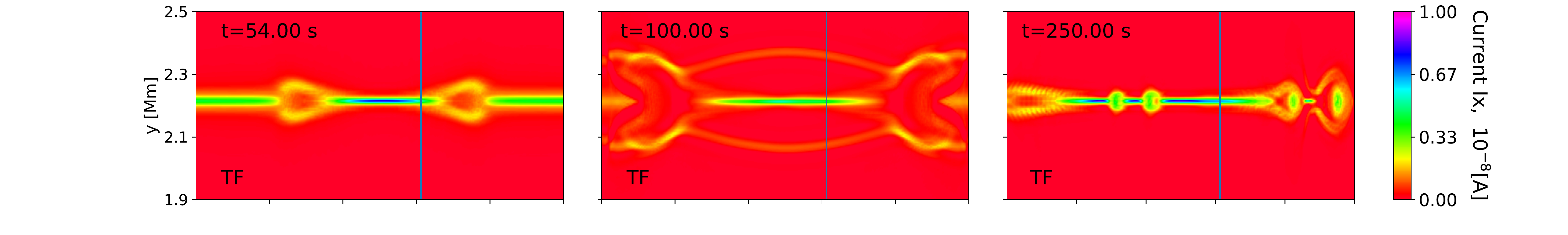}\vspace*{-0.2cm}\\
	\hspace*{-1.7cm}
	\includegraphics[scale=0.4]{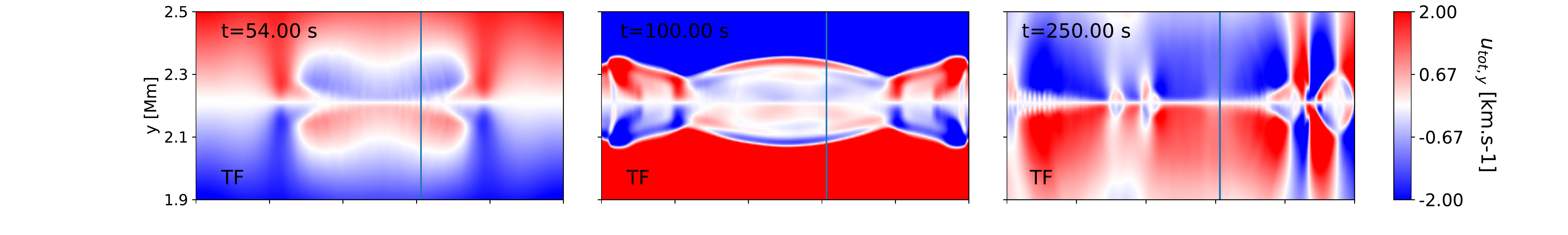}\vspace*{-0.2cm}\\
	\hspace*{-1.7cm}
	\includegraphics[scale=0.4]{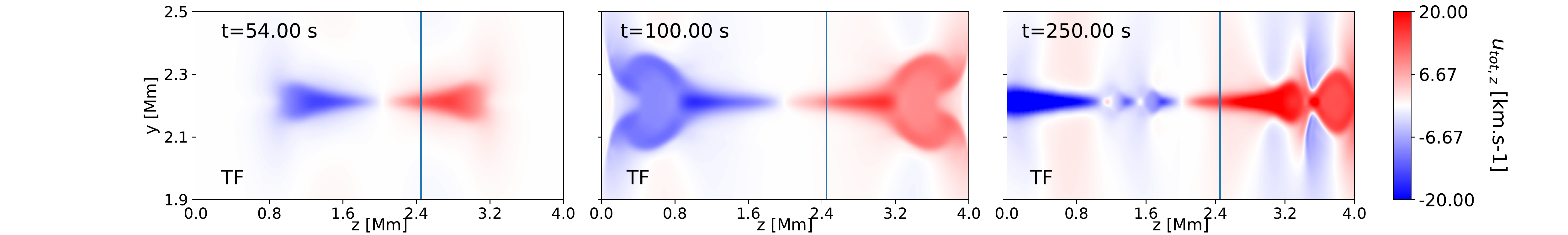}\vspace*{-0.2cm}
		\caption{Distribution of the component $x$ of the current density $\totalcurrent_x$ in Amp\`ere and total velocity in km s$^{-1}$ in the $y$ and $z$ directions corresponding to the inflow and outflow directions of the reconnection region respectively, for the MFMS case (first three rows) and TF case (last three rows) in a restricted domain defined as $(y,z)\in[1.9, 2.5]\times[0,4]$~Mm. From left to right: evolution of these distributions at $t=54,\, 100$ and $250$~s. The vertical line corresponds to $z=2.45$~Mm. The associated animation of this figure shows the time evolution.}
		\label{fig:current}
\end{figure*}

Figure~\ref{fig:recorate} shows the evolution of the thickness of the current sheet at a fixed $z=2.45$~Mm (top left), the mean thickness of the whole current sheet (top right), the length of the current sheet (bottom left), and the maximum total velocity in the $z$ direction $u_{\text{tot},z}$ (bottom right) taken at the main reconnection site, for both MFMS and TF. To calculate the length at a fixed $z$, first, we locate the maximum of the magnitude of current for any $y\in[2.1,\,2.3]$~Mm. Then, we consider its full-width half maximum (FWHM) in the $z$ direction to estimate the length. In our simulations, the maximum current is generally located at $y=L_0/2$. Similarly, the $z$ component of the maximum total velocity at the reconnection site has been taken within the FWHM along the $z$ axis of the maximum current. Concerning the thickness at a fixed $z=2.45$~Mm, a similar procedure as the estimation of the length of the current sheet has been performed. First, we locate the maximum magnitude of the current for any $y$ in the domain at a fixed $z=2.45$~Mm. We choose the FWHM as it accurately represents the termination of the reconnection site for both the laminar and plasmoid instability phases. We extend these calculations to any $z\in[0.8,3.2]$~Mm in order to calculate the evolution of the mean thickness and deviation of the current sheet. 

Focusing on the left column of Fig.~\ref{fig:recorate}, from 0 to approximately $25$~s, for both MFMS and TF, the thickness at $z=2.45$~Mm and length of the current sheet decreases drastically. It corresponds to a fast transition from the initial conditions to the beginning of the laminar phase leading to the formation of the current sheet. The initial thickness of the current sheet is $\lambda_{\phi}=0.1 L_0$ and the length $L_0 = 426$~km. In both MFMS and TF, from $0$\, to $\sim 100$~s, the thickness at $z=2.45$~Mm of the current sheet decreases to  $\sim 25$~km corresponding to the numerical length scale. Additionally, if we focus on the mean thickness of the current sheet (top right of Fig.~\ref{fig:recorate}), it is clear that the thickness is larger in MFMS than in TF during the laminar phase. Consequently, TF reduces the thickness of the current sheet slightly faster than in MFMS, leading to higher velocities during the laminar regime, as shown in the bottom right of Fig.~\ref{fig:recorate}. Note that there are no large disparities between MFMS and TF in the evolution of the length of the current sheet.

At $t\approx100~s$, the plasmoid instability phase of the reconnection process starts. This phase is characterized by strong variation in the thickness of the current sheet as shown in top left plot of Fig.~\ref{fig:recorate}. In MFMS, we can clearly see that the thickness at $z=2.45$~Mm increases around at $t=200$~s from 25 to $80$~km. Similarly in TF, a small variation of the thickness at $z=2.45$~Mm is obtained around $t=140$~s where the thickness increases from 25 to approximately $37$~km.
In general, at any time, the average size of the plasmoids appears to be slightly larger in MFMS than in TF meaning that the dilatation of the plasmoids during their formation is more important in MFMS than in TF, as shown in the top right panel of Fig.~\ref{fig:recorate}. 

The plasmoid instability phase impacts the length of the current sheet that is associated with the maximum current $\totalcurrent_x$ in both MFMS and TF (bottom left plot of Fig.~\ref{fig:recorate}). Indeed, after $t\approx100$~s, in both MFMS and TF, strong variations of the length of the current sheet are obtained, but on average decrease. This is because the presence of plasmoids breaks the initial current sheet formed during the laminar phase into multiple thinner current sheets. In summary, even during the plasmoid instability phase, on average, the mean frequency at which the plasmoids are formed and length scales between them is similar between MFMS and TF.

At any time of the reconnection process, in both MFMS and TF, the maximum total velocity in the $z$ direction $u_{\text{tot},z}$ is increasing, as illustrated in the bottom right of Fig.~\ref{fig:recorate}. However, it is clear that the velocity is  higher in TF than in MFMS, meaning that stronger flows are obtained in TF than in MFMS, in both the laminar and plasmoid instability phase. Note that a similar result has been obtained in the $y$ direction.

\begin{figure*}[tbh]
	\begin{center}
	   \hspace*{-0.4cm}
		\includegraphics[scale=0.35]{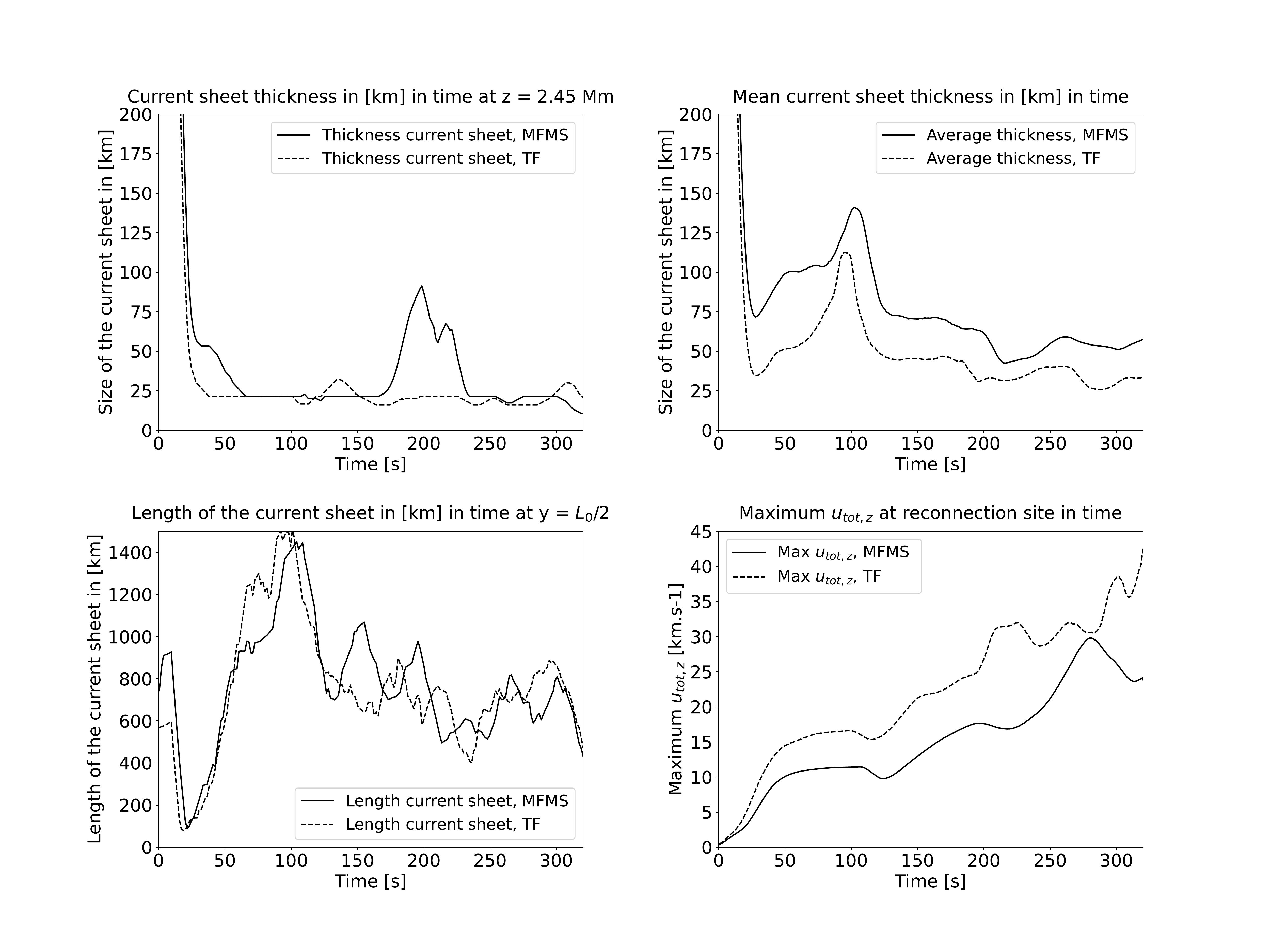}
		\vspace*{-0.7cm}
		\caption{Evolution in time of the thickness at a fixed $z=2.45$~Mm (top left), mean thickness for $y\in[1,\,3.5]$~Mm (top right) and length of the current sheet (bottom left), and maximum total velocity in the $z$ direction estimated at the ouflow of the main reconnection site (bottom right), by taking the full with half maximum (FWMH) of the maximum current density $\totalcurrent_x$. The dashed and full line corresponds to TF and MFMS respectively. Note that $z=2.45$~Mm is illustrated in Fig.~\ref{fig:current} by the black vertical line.}
		\label{fig:recorate}
	\end{center}
\end{figure*}

In summary, the average thickness of the current sheet and the size of the plasmoids appears to be slightly larger in MFMS than in TF, and, reconnection with the TF model develop stronger flows than the MFMS model. In this context, the presence of helium species appears to only slightly impact the evolution of the current sheet as well as the flows. The length of the current sheet and the frequency at which the plasmoids are formed are similar between the two models. In the next sections, we will see that, even though these properties only slightly differ between the two models, the decoupling of the particles and composition of the plasma in the current sheet show large disparities, leading to different evolution of the heating processes and temperatures.

\subsection{Decoupling of the particles and velocity fields}
\label{sec:decoupling}

In this section, the decoupling of the particles and velocity fields are investigated. To do so, we focus on the $z$ and $y$ components of the velocity for each fluid separately. These correspond to the outflow and inflow directions of the magnetic reconnection, respectively. 


The first four rows (starting from the top) of Figures~\ref{fig:2dvelocityy} (Fig.~\ref{fig:2dvelocityz}) represent the distribution of the $y$ ($z$) component of the velocity of protons H$^+$ and the drift velocities with all the other particles in MFMS. The last two rows of the figures  correspond to the velocity of protons H$^+$ and drift velocity between H and H$^+$ in TF. From left to right columns, it shows the time evolution at times $t=89$~s,\, $t=282$~s and $t=327.5$~s, respectively.

\begin{figure*}[tbh]
	\begin{center}
	\hspace*{-1.7cm}
	\includegraphics[scale=0.4]{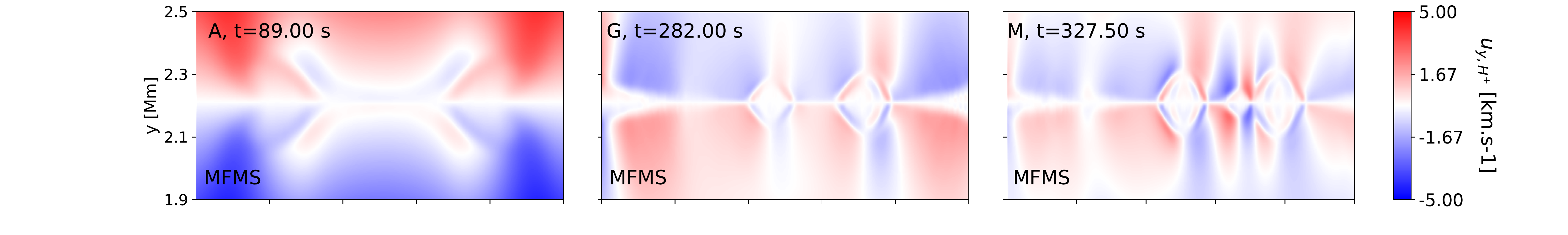}\vspace*{-0.2cm}\\
	\hspace*{-1.7cm}
	\includegraphics[scale=0.4]{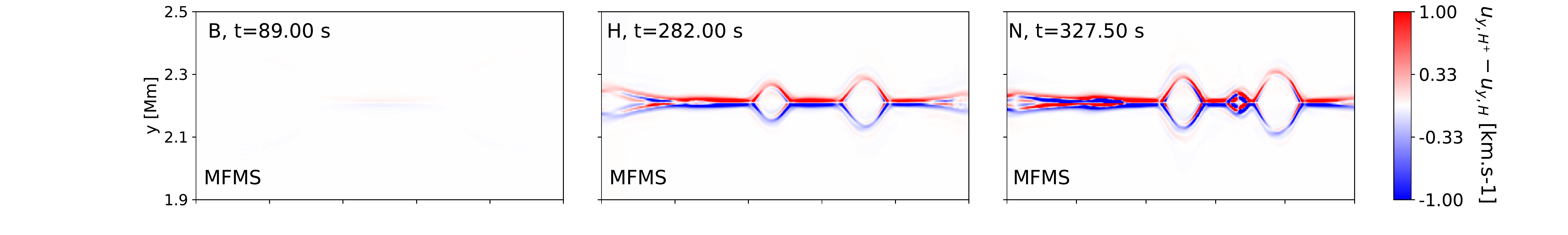}\vspace*{-0.2cm}\\
	\hspace*{-1.7cm}
	\includegraphics[scale=0.4]{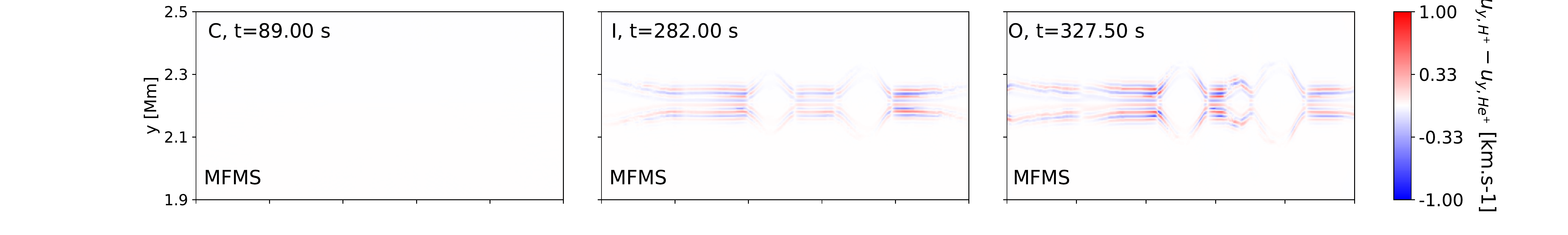}\vspace*{-0.2cm}\\
	\hspace*{-1.7cm}
	\includegraphics[scale=0.4]{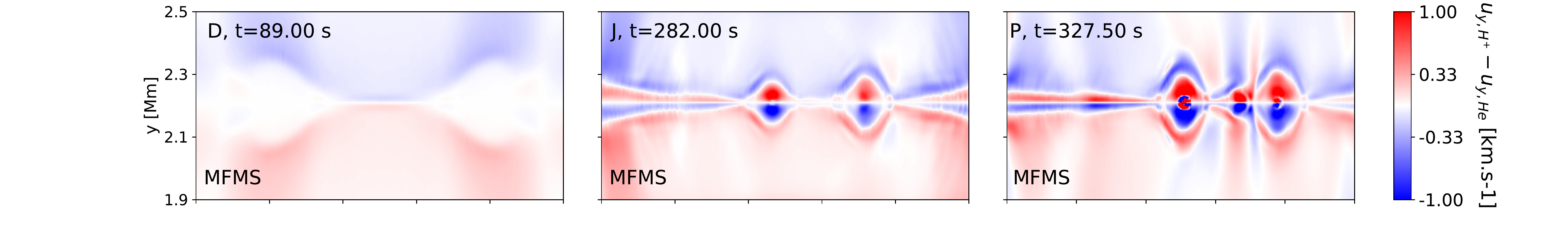}\vspace*{-0.2cm}\\
	\hspace*{-1.7cm}
	\includegraphics[scale=0.4]{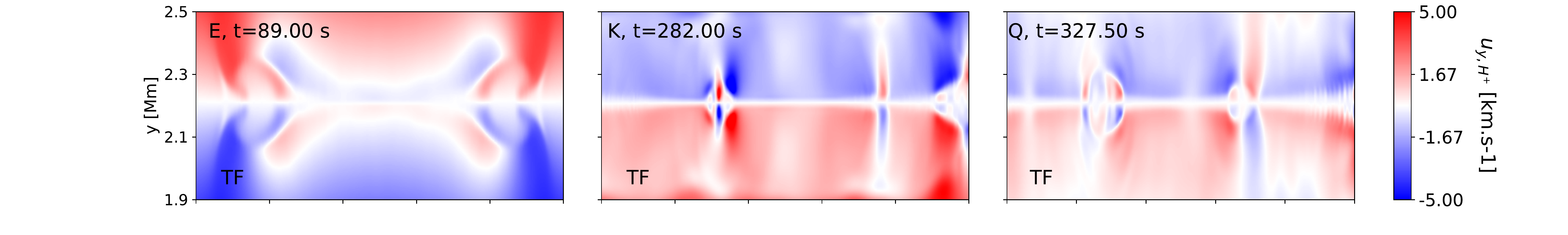}\vspace*{-0.2cm}\\
	\hspace*{-1.7cm}
	\includegraphics[scale=0.4]{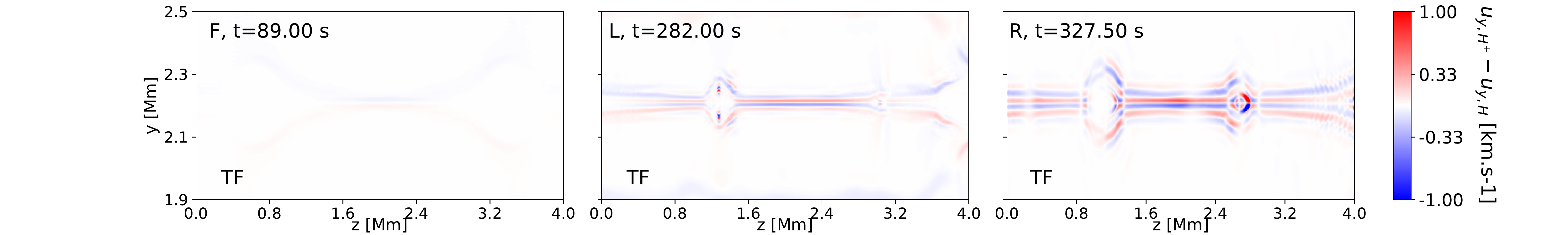}\vspace*{-0.2cm}
		\caption{Distribution of the $y$ component of the velocity of H$^+$ and drift velocities with other species in km\,~s$^{-1}$ at different times for MFMS and TF. From first to fourth row: distribution of the velocity $u_{y,\text{H}^+}$, the drift velocities $u_{y,\text{H}^+}-u_{y,\text{H}}$, $u_{y,\text{H}^+}-u_{y,\text{He}^+}$ and $u_{y,\text{H}^+}-u_{y,\text{He}}$ in MFMS respectively. The last two rows corresponds to the distribution of the velocity $u_{y,\text{H}^+}$ and the drift velocity $u_{y,\text{H}^+}-u_{y,\text{H}}$ in TF. From left to right column: different times $t=136$,\, $t=278.5$ and $327.5$~s are presented. An animation of this figure is available.}
		\label{fig:2dvelocityy}
	\end{center}
\end{figure*}

\begin{figure*}[tbh]
	\begin{center}
 		\hspace*{-1.7cm}
	\includegraphics[scale=0.4]{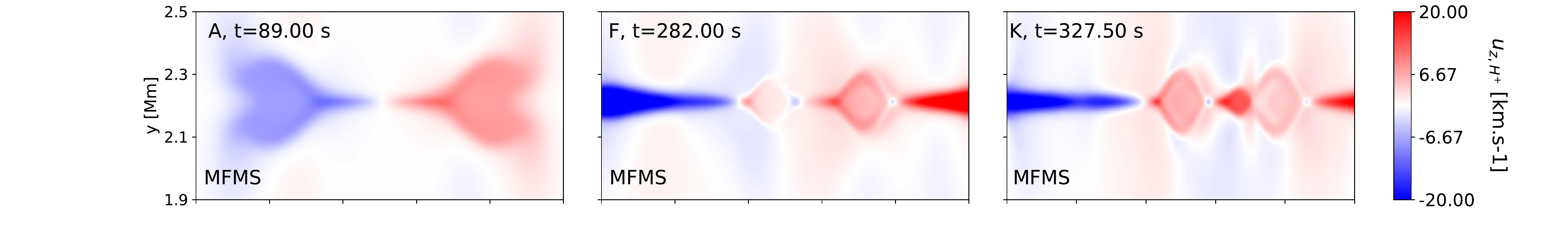}\vspace*{-0.2cm}\\
	\hspace*{-1.7cm}
	\includegraphics[scale=0.4]{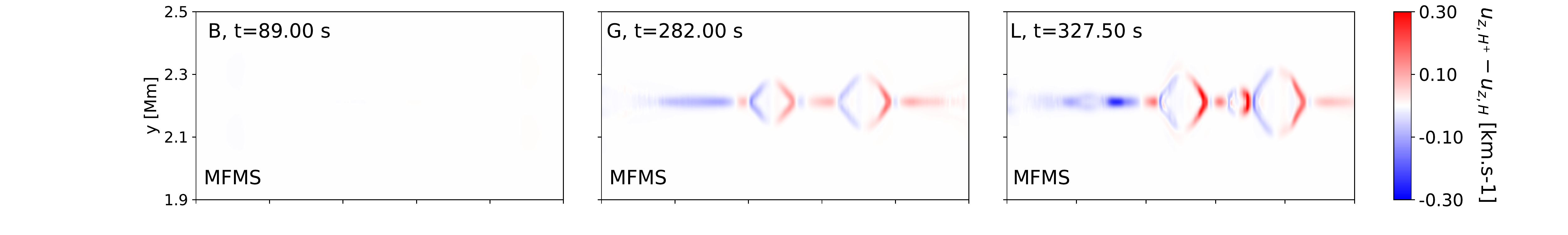}\vspace*{-0.2cm}\\
	\hspace*{-1.7cm}
	\includegraphics[scale=0.4]{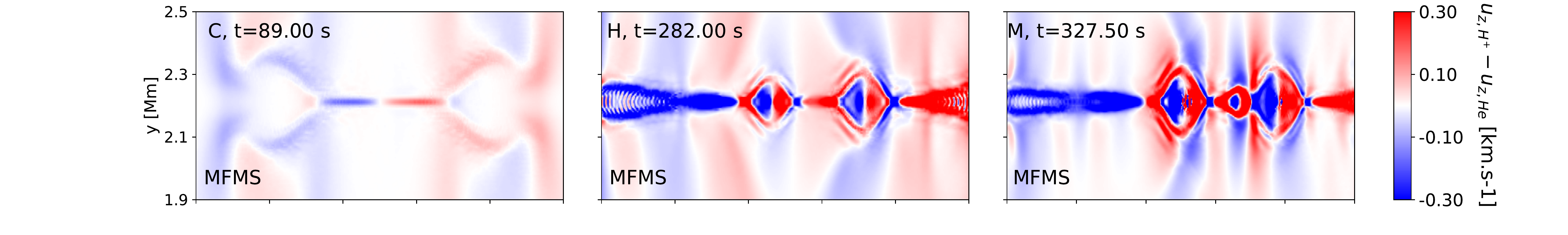}\vspace*{-0.2cm}\\
	\hspace*{-1.7cm}
	\includegraphics[scale=0.4]{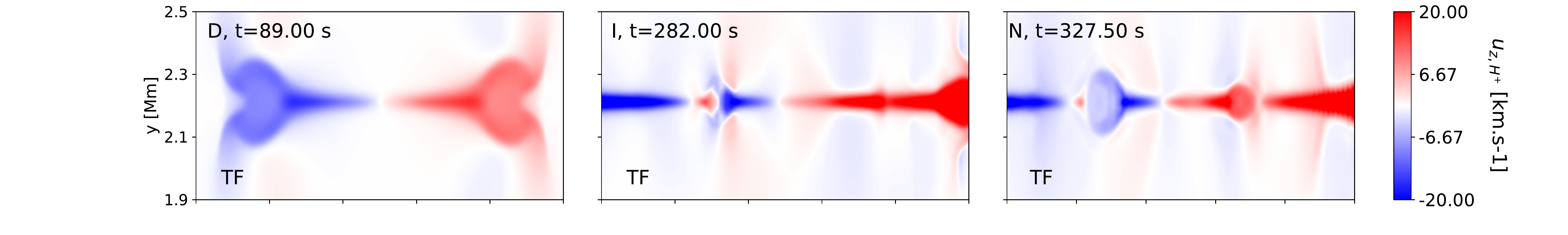}\vspace*{-0.2cm}\\
	\hspace*{-1.7cm}
	\includegraphics[scale=0.4]{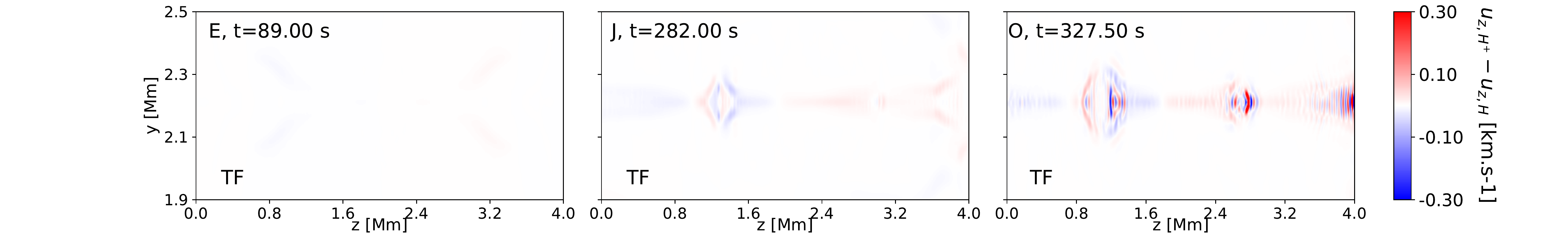}\vspace*{-0.2cm}
		\caption{Similar layout as Fig.~\ref{fig:2dvelocityy} for the $z$ component of the velocities. Unlike Fig.~\ref{fig:2dvelocityy}, the drift velocity between H$^+$ and He$^+$ is not shown since its magnitude is negligible compared to the other quantities. An animation of this figure is available.}
		\label{fig:2dvelocityz}
	\end{center}
\end{figure*}


During the laminar phase, the velocities of H$^+$ increases up to a maximum achieved at approximately $1$~km s$^{-1}$ and 20~km s$^{-1}$ $y$ and $z$ components in both simulations. Note that TF seems to show slightly higher velocities, as described in Section \ref{sec:currentsheet}. This maximum is reached at the end of the laminar phase which corresponds to $t\approx100$~s for both MFMS and TF. This corresponds to the time where the maximum length of the current sheet is achieved, as shown in Fig.~\ref{fig:recorate}. In the first column of Figs.~\ref{fig:2dvelocityy}, and ~\ref{fig:2dvelocityz}, the drift velocities between particles are small compared to $u_{y,\text{H}^+}$ and $u_{z,\text{H}^+}$ which implies that there is a large collisional coupling between species during the laminar phase. The decoupling between H$^+$ and He is highest, as shown in panel D of Fig.~\ref{fig:2dvelocityy} and F of Fig.~\ref{fig:2dvelocityz}. Neutral helium species moves slightly slower  than H$^+$ by approximately $0.25$ and 0.15~km s$^{-1}$ in the y and z directions respectively. Concerning TF, no significant decoupling between H and H$^+$ is seen.

In order to understand the evolution of the velocity fields, we have also calculated and analyzed the distributions of all the forces acting on each fluid involved in MFMS or TF, at different times. In the following, for the sake of simplicity, we briefly summarize them.

During the laminar phase, the initial structure of the magnetic field allows the development of a Lorentz force and leads to the formation of outflow and inflow velocities. The Lorentz force and gradient of pressure associated with H$^+$ are the dominant forces of the plasma in both MFMS and TF. Therefore, during the laminar phase, the dynamics of the reconnection process itself in the current sheet are dominated by these forces. This is also consistent with the fact that H$^+$ is the most dominant species in number and mass density in both MFMS and TF at that time. In MFMS, no drift velocities associated with H$^+$-H and H$^+$-He$^+$ occur since the species H$^+$, H and He$^+$ are highly coupled due to collisions. In other words, in each individual momentum equation associated with H and He$^+$, the magnitude of the momentum exchange terms due to collisions with H$^+$ is large enough to highly couple the momentum equations associated with H, H$^+$ and He$^+$ altogether. Similarly in TF, the fluids H and H$^+ $ are highly coupled due to collisions during the laminar phase.

In contrast, in MFMS, the neutral helium species is decoupled from the other fluids. This drift at the current sheet is due to the fact that neutral He decouple from H$^+$ when the thickness of the current sheet thins down to the order of the mean free path between He and H$^+$ during the laminar phase, and, neutral He are not dragged by collisions into the current sheet as quickly as H$^+$ or H. The decoupling between He and H$^+$ due to the lack of collisions is clearly illustrated in panel D of Fig.~\ref{fig:2dvelocityy} and in panel C of Fig.~\ref{fig:2dvelocityz}. The fact that the decoupling between He and H$^+$ is stronger than other interactions is attributed to the collisional frequencies that involves neutral helium species which are much smaller by more than two orders of magnitude than all the other collisional frequencies, as shown in Table ~\ref{tab:table_colfreqHP} (for the initial conditions). These differences are attributed to the magnitude of the collision integrals at the kinetic level that leads to smaller collisional frequencies, as described with further details in \citet{Wargnier2022}. This effect leads to a slower velocities for neutral helium species with respect to H$^+$, as illustrated by panel D of Fig.~\ref{fig:2dvelocityy} and panel C of Fig.~\ref{fig:2dvelocityz}. 


In the second and last column of Figs.~\ref{fig:2dvelocityy} and ~\ref{fig:2dvelocityz} corresponding to the plasmoid instability phase, we note that drift velocities are increasing during the formation of plasmoids for both MFMS and TF. This is also illustrated in the animations. When plasmoids are produced, local velocity fields in opposite directions are generated due to a local reconnection of the magnetic field lines. It should be noted that the structure and evolution of the velocity of H$^+$ is similar for both MFMS and TF, however, the drift velocities differ between the two. In the case of MFMS, the drift between H and H$^+$ is stronger at the boundary of plasmoids and inside the current sheet, reaching approximately 1 and 0.1~km s$^{-1}$ for the y and z components respectively, as we can see in panels H and N of Fig.~\ref{fig:2dvelocityy} and panels G and L of Fig.~\ref{fig:2dvelocityz}. In TF, the drift velocity between H and H$^+$ is smaller than in MFMS by a factor of three and is showing some oscillating pattern structure parallel to the current sheet, as shown in panel R of Fig.~\ref{fig:2dvelocityy}. Similar to the decoupling between He and H$^+$, the drift between H and H$^+$ develops during the plasmoid instability phase, as the current sheet thins down to the order of the mean free path (a few kilometers corresponding to a few grid points) between H and H$^+$ and neutral H does not collide enough with H$^+$ to be dragged into the current sheet as quickly as H$^+$. This decoupling is consistent with the results obtained by \citet{Leake:2012pr,Leake:2013dq}. 


The drift between H and H$^+$ is higher in MFMS than in TF because the collisional frequency associated with H-H$^+$ interactions is smaller in MFMS than in TF. As shown in \citet{Wargnier2022}, the collisional frequencies are calculated from the products of 1- the collision integral associated with H-H$^+$ interaction, 2- H and H$^+$ number densities and 3- thermal speed associated with H-H$^+$ interaction. In our chosen conditions, the number densities of H and H$^+$ in MFMS are smaller than in TF due to the presence of helium species. Additionally, the temperature inside the current sheet is higher in MFMS than in TF, as shown later in Section~\ref{sec:heating}, leading to a decrease of the collision integral associated with H-H$^+$ interactions, as shown in \citet{Wargnier2022}. These two effects contribute to decrease the collisional frequency associated with H-H$^+$ interactions in MFMS compared with TF. Note that if the temperature increases, the thermal speed contributes to increase the collisional frequency, as shown in the definition from \citet{Wargnier2022}. However, in the chosen conditions and for H-H$^+$ interactions, this contribution is smaller than the contribution of the H-H$^+$ collision integral and number densities. 


In panels L and R of Figure~\ref{fig:2dvelocityy} and panels J and O of Figure~\ref{fig:2dvelocityz}, the oscillating pattern of the drift velocity between H and H$^+$ in TF is attributed to the spatial distribution momentum exchange term associated with H-H$^+$ interactions. These collisions compensate for the localized unbalance of the forces \ie\, the Lorentz force and pressure gradients, when the current sheet thins down to the mean free path between H and H$^+$. A similar effect has been obtained in the distribution of drift velocity between H$^+$ and He$^+$ in panel O of Fig.~\ref{fig:2dvelocityy} where the characteristic wavelength and magnitude of this pattern are much smaller than for the drift between H and H$^+$. However, this effect is attributed to two phenomena: the momentum exchange term associated with H$^+$-He$^+$ interactions in MFMS tries to compensate for the localized unbalance of forces (pressure gradients of H, H$^+$, He and He$^+$ and Lorentz forces acting on H$^+$ and He$^+$ separately) at much smaller spatio-temporal scales than in TF. In  addition, the ions experience a synchronized cyclotron-type motion along the magnetic field lines, also known as the ion-coupling effect, leading to a sinusoidal ion drift velocity distribution perpendicular to the magnetic field line, as described with further details in \citet{Martinez_Sykora_2020}. Note that the $z$ component of the drift velocity associated with H$^+$-He$^+$ interactions is negligible. Consequently, the distribution of the latter has not been shown here.
 
As in the laminar phase, the decoupling between He and H$^+$ remains important due to the lack of collisions between the two species. This drift velocity is high both inside the current sheet and plasmoids. The maximum drift velocity between He and H$^+$ reaches more than 1~km s$^{-1}$ and 0.3~km s$^{-1}$ in the $y$ and $z$ directions respectively.  We can also notice some internal structure of the drift velocity inside the plasmoids (panels J and P of Fig.~\ref{fig:2dvelocityy} and panels H and M of  Fig.~\ref{fig:2dvelocityz}).

\subsection{Multi-fluid heating processes}
\label{sec:heating}

\subsubsection{Heating processes due to departure in velocities}
\label{sec:heatingdrift}

In order to analyze the heating processes and identify the type of collisions mainly responsible for the increase in temperature, we focus on the dominant heating terms due to drift velocities $Q_{\specie\ions,\specie'\ions'}^{\textbf{u},\text{col}}$ in both models MFMS and TF. As explained in Section~\ref{sec:ener}, the heating terms associated with drift velocities are not adiabatic terms and can increase the total thermal energy of the plasma. Therefore, the heating processes for both MFMS and TF could be strongly connected to the structure of the drift velocities which have been investigated in previous Sections~\ref{sec:decoupling}. 

Under the chosen thermodynamic conditions presented in Section~\ref{sec:methodology}, the electron thermal conduction and the terms associated with ionization or recombination processes $Q_{\specie\ions}^{\text{ion}}$ or $Q_{\specie\ions}^{\text{rec}}$ have negligible impact on the heating processes, in both MFMS and TF. Indeed, as described in Section~\ref{sec:model}, the heating terms associated with ionization or recombination processes for any $\specie\ions\in\heavy$ do not contribute to the total thermal energy of the plasma (apart from radiative losses term). Thus, they do not contribute to the current sheet's temperature increase. In all thermal energy equations, the magnitude of these terms is negligible compared to the thermalization terms due to collisions since ionization and recombination timescales are bigger than collisional timescales, as described in Section~\ref{sec:composition} and also in Section~\ref{sec:ionization}.

Figure~\ref{fig:heating_laminar} represents the distribution of the rate of the heating terms $Q_{\specie\ions,\specie'\ions'}^{\textbf{u},\text{col}}$ in a 1D cut of the current sheet located at $z=2.45$~Mm for all the possible interactions involved in both MFMS (top row) and TF (middle row) at different times $t=11.5$~s and $t=28.5$~s from left to right column. Note that these times correspond to a different stage of the laminar phase. The rates are calculated by dividing the heating terms, produced by drift velocities, by the total thermal energy $e^{\alpha}_{\text{tot}}=\sum_{\specie\ions\in\alpha}e_{\specie\ions},\,\alpha\in\{\mixture, \mixture_{\text{TF}}\}$. We also show the distribution of the temperatures of each species in MFMS and TF (bottom row) at $z=2.45$~Mm. Note that an animation of this figure is available.

At the beginning of the laminar phase in MFMS, it is clear that the heating processes produced by the drifts between H$^+$ and He, H and He dominate in the current sheet, as observed in panel A of Fig.~\ref{fig:heating_laminar}. This effect is consistent with the previous Sections~\ref{sec:decoupling} showing that the drift velocities between neutral helium and hydrogen species are much higher than any other interactions during the laminar phase. Additionally, the heating due to collisions between electrons and all heavy species (ohmic collisions) is at the same order of magnitude as H$^+$-He and H-He interactions but localized around the current sheet. The heating due to the drift between H$^+$ and H is not negligible in the current sheet. Still, it is much lower than the other interactions (panel A of Fig.~\ref{fig:heating_laminar}), which is also consistent with the magnitude of the drift velocity in Sections~\ref{sec:decoupling}. In the case of TF, the heating produced by the drift between electrons and heavy species (H and H$^+$) is dominating, as shown in panel B. Additionally, the latter is slightly higher in magnitude as the heating is produced by the drift between electrons and heavy species in panel A (black dashed line in panel A). However, if we sum all the heating terms of panel A and compare it with the sum of all the heating terms in panel B, it is clear that the total heating produced by velocity drifts is higher, by approximately a factor two, in MFMS than TF due to the heating terms $Q_{\text{H},\text{He}}^{\textbf{u},\text{col}}+Q_{\text{He},\text{H}}^{\textbf{u},\text{col}}$ and $Q_{\text{H}^+,\text{He}}^{\textbf{u},\text{col}}+Q_{\text{He},\text{H}^{+}}^{\textbf{u},\text{col}}$.

At a later time of the laminar phase, corresponding to the second column of Fig.~\ref{fig:heating_laminar}, the temperatures of all species in MFMS increase and are higher than in TF. The temperature achieves its maximum at the center of the current sheet at $27\,000$~K in MFMS. In contrast, in TF, the temperature is approximately $19\,000$~K at the center of the current sheet (panel F). By comparing panels A and D, the magnitude of the heating terms due to collisions associated with interactions H-He, H$^+$-He, and electrons-heavy species increase in time inside the current sheet. This increase is attributed to drift velocities increasing with time during the laminar phase. Similarly, if we sum all the heating terms in panels A and B, the sum of all the heating terms is higher in MFMS than in TF. Consequently, temperatures inside the current sheet are higher in MFMS than in TF. Note that temperatures of all species are equal during the laminar phase in both MFMS and TF.

\begin{figure*}[tbh]
	\begin{center}
	\includegraphics[scale=0.32]{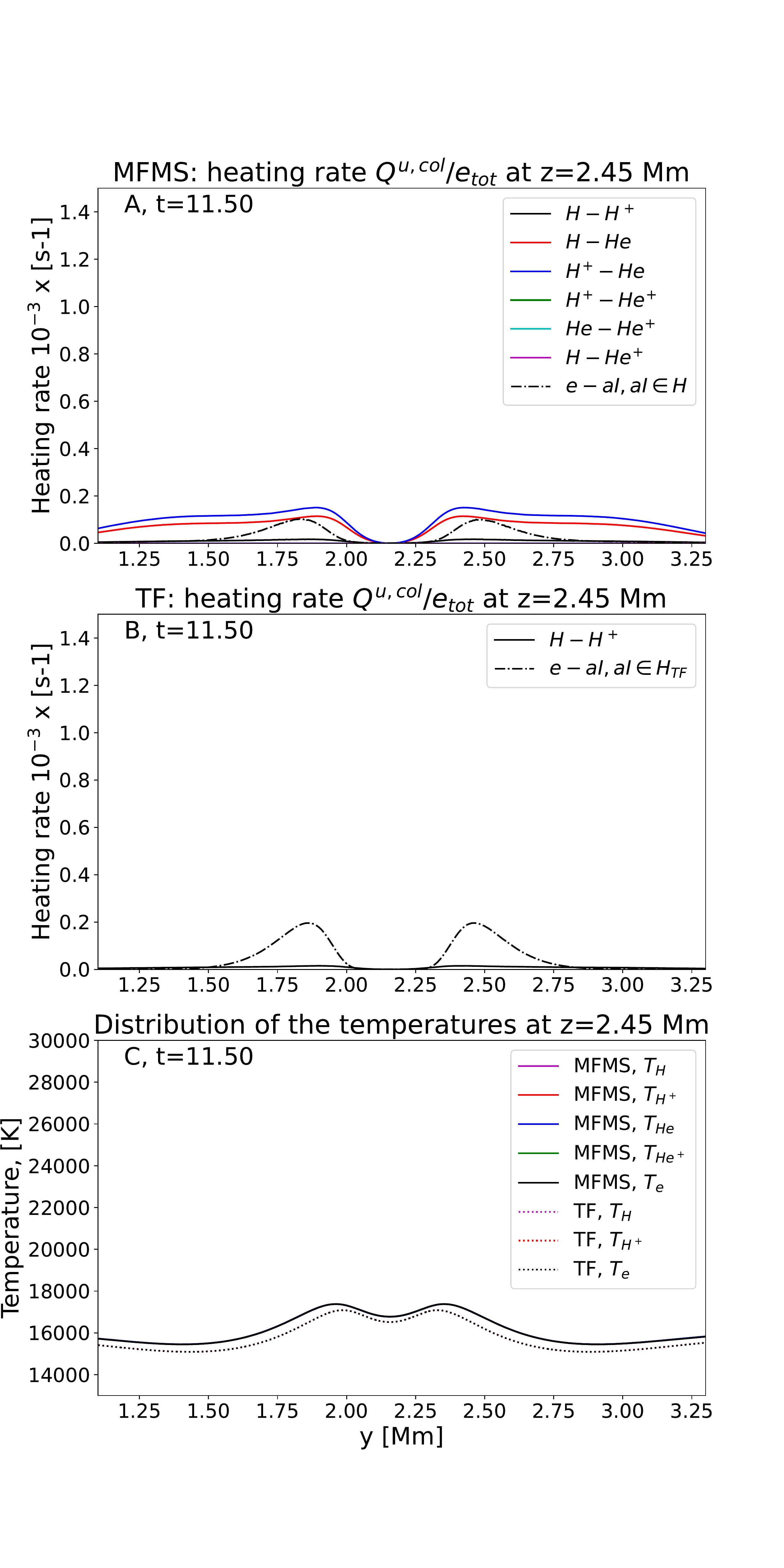}\vspace*{-0.2cm}
	\hspace*{-0.1cm}
	\includegraphics[scale=0.32]{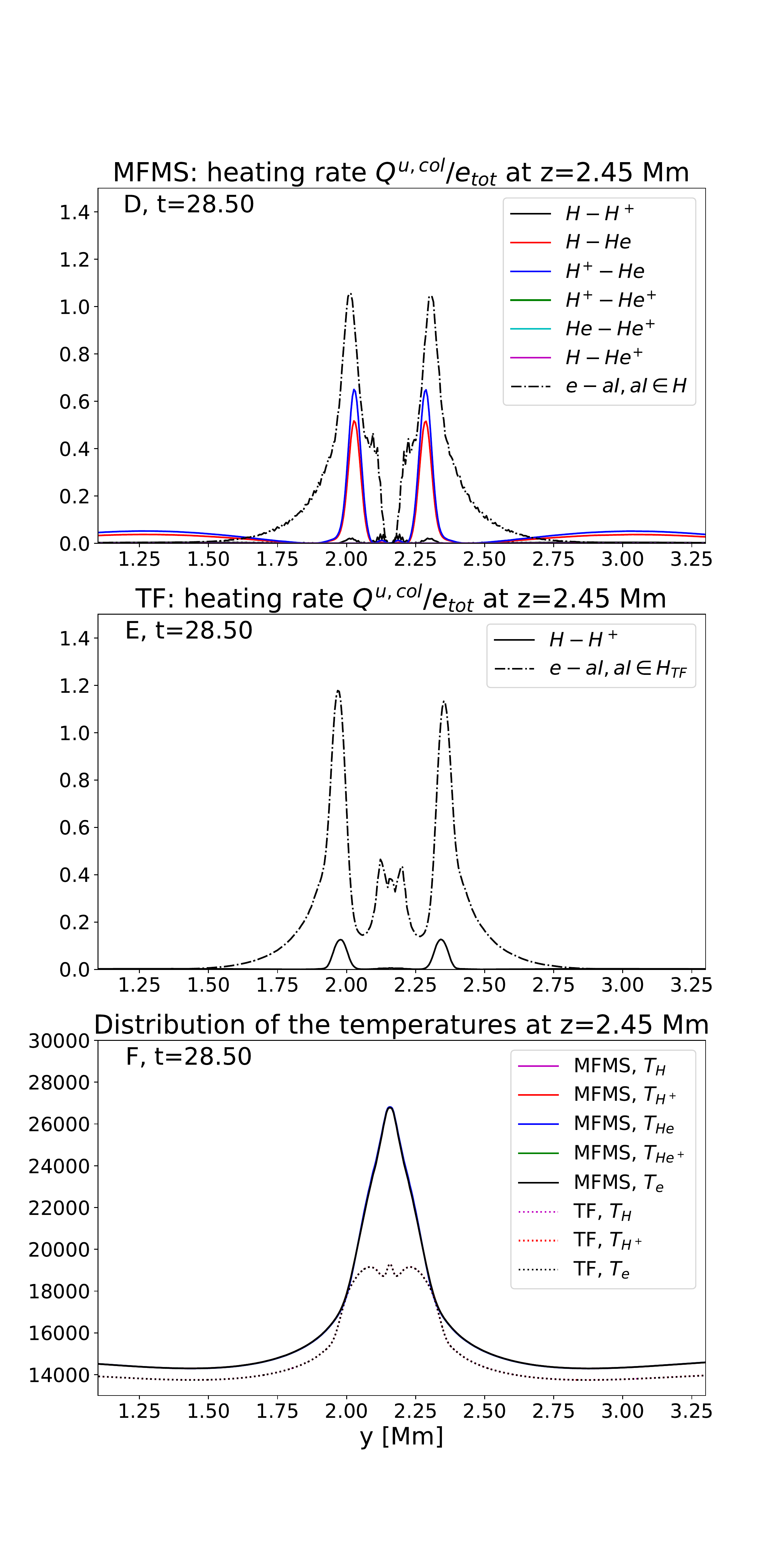}\vspace*{-1cm}
		\caption{Distribution of the rates of the heating terms $Q_{\specie\ions,\specie'\ions'}^{\velocity,\text{col}}+Q_{\specie'\ions',\specie\ions}^{\velocity,\text{col}},\,(\specie\ions,\specie'\ions')\in\mixture^2$ or $\mixture^2_{\text{TF}}$ associated with drift velocities divided by the total thermal energy $e_{\text{tot}}=\sum_{\specie\ions\in\mixture}e_{\specie\ions}$ in s$^{-1}$ (top row for MFMS and middle row for TF) and temperatures of each particles (bottom row) along the $z$ axis in the current sheet at position $y=L_0/2$ in both MFMS and TF, respectively. From left to right column: distributions at different times $t=11.5$~s and $t=28.5$~s corresponding to the laminar phase. Note that the vertical axis ranges from $0$ to $1.4\times 10 ^{-3}$~s$^{-1}$. An animation of this figure is available.}
		\label{fig:heating_laminar}
	\end{center}
\end{figure*}

\begin{figure*}[tbh]
	\begin{center}
       \hspace*{-1.7cm}
	\includegraphics[scale=0.4]{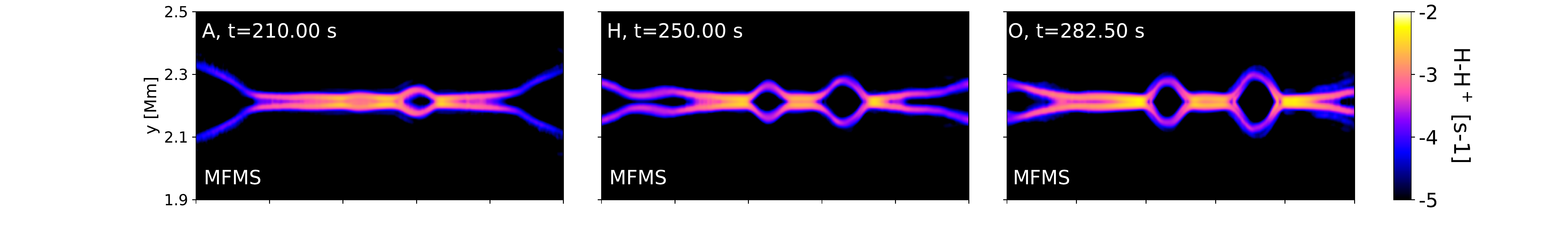}\vspace*{0.cm}\\
	\hspace*{-1.7cm}
	\includegraphics[scale=0.4]{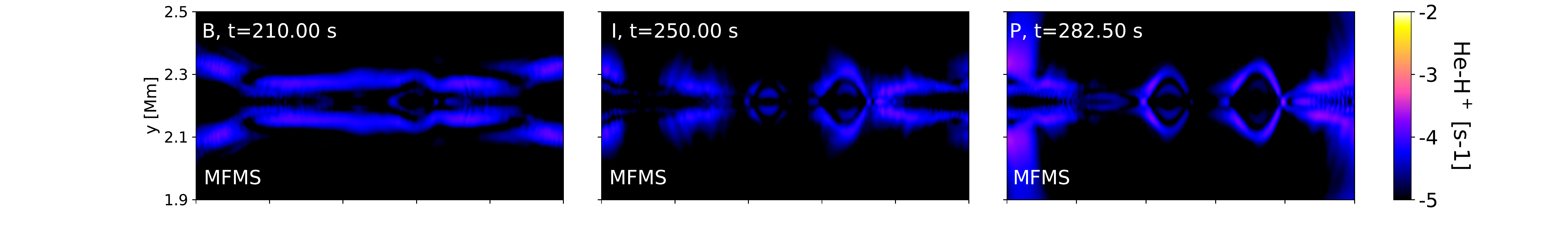}\vspace*{-0.2cm}\\
	\hspace*{-1.7cm}
	\includegraphics[scale=0.4]{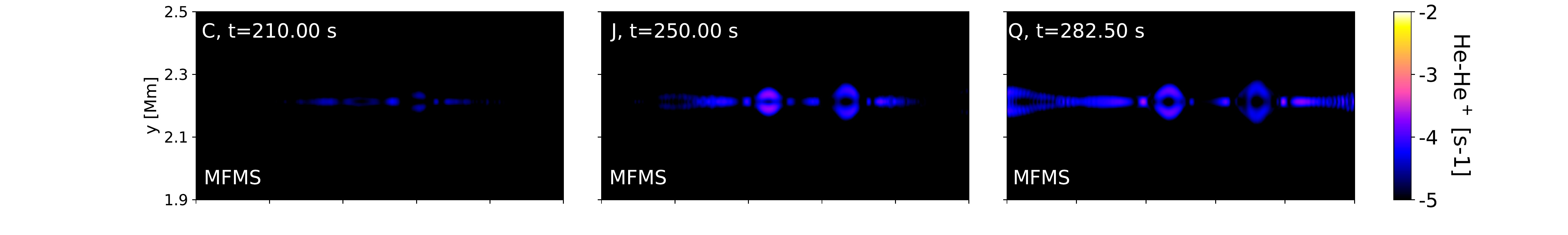}\vspace*{-0.2cm}\\
	\hspace*{-1.7cm}
	\includegraphics[scale=0.4]{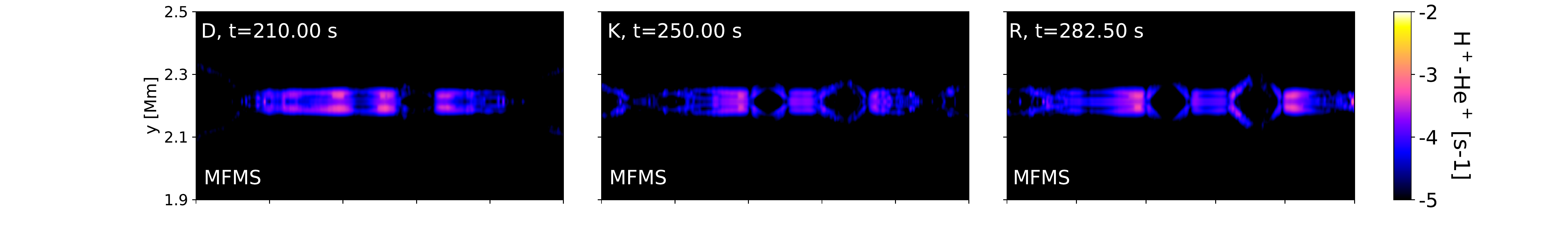}\vspace*{-0.2cm}\\
	\hspace*{-1.7cm}
	\includegraphics[scale=0.4]{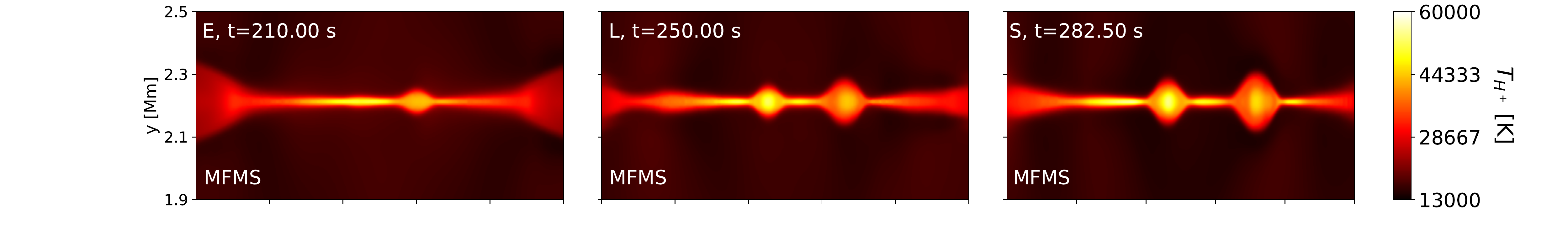}\vspace*{-0.2cm}\\
	\hspace*{-1.7cm}
	\includegraphics[scale=0.4]{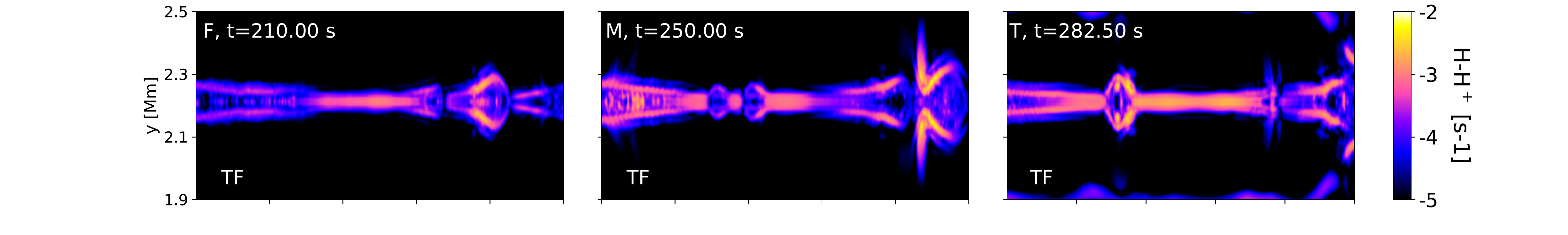}\vspace*{-0.2cm}\\
	\hspace*{-1.7cm}
	\includegraphics[scale=0.4]{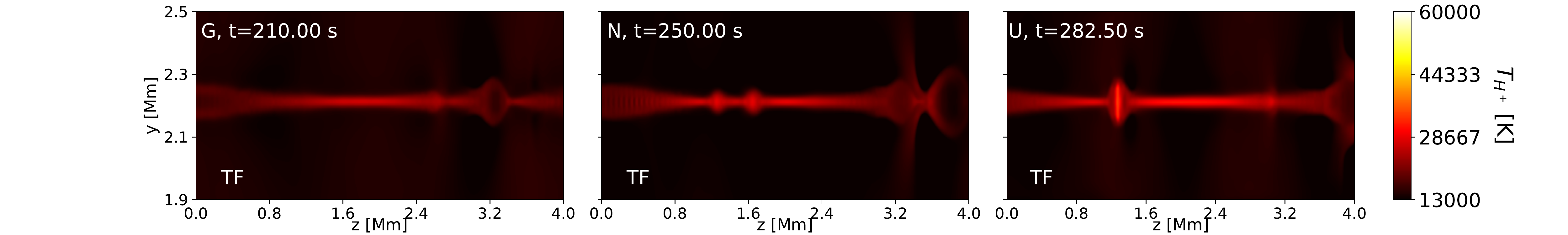}\vspace*{-0.2cm}
		\caption{From top to bottom, for any $(y,z)\in[1.9,2.5]\times[0,4]$~Mm: distribution of the decimal logarithm of the rates of the heating terms $Q_{\specie\ions,\specie'\ions'}^{\velocity,\text{col},S},\,(\specie\ions,\specie'\ions')\in\mixture^2$ divided by the total thermal energy $e_{\text{tot}}=\sum_{\specie\ions\in\mixture}e_{\specie\ions}$ in s$^{-1}$ in MFMS, the temperature of H$^+$ in MFMS, the decimal logarithm of the heating terms $Q_{\specie\ions,\specie'\ions'}^{\text{col}},\,(\specie\ions,\specie'\ions')\in\mixture_{\text{TF}}^2$ divided by the total thermal energy $e_{\text{tot}}=\sum_{\specie\ions\in\mixture_{\text{TF}}}e_{\specie\ions}$ in s$^{-1}$ in TF and the temperature of H$^+$ in TF. The other heating terms have not been shown here since their magnitude are negligible compared to the other terms. From left to right: distributions at different times $t=210$, $250$ and $282.5$~s corresponding to different stages of the plasmoid instability phase. An animation of this figure is available.}
		\label{fig:heating}
	\end{center}
\end{figure*}
Figure~\ref{fig:heating} shows the heating rates due to the drift velocities $Q_{\specie\ions,\specie'\ions'}^{\velocity,\text{col}}+Q_{\specie'\ions',\specie\ions}^{\velocity,\text{col}},\,(\specie\ions,\specie'\ions')\in\mixture^2$ or $\mixture_{\text{TF}}^2$ divided by the total thermal energy and the temperature of protons $T_{H^+}$, for MFMS (first five rows) and TF (last two rows), at different times of the simulation corresponding to the plasmoid instability phase. From left to right columns, it corresponds to times $t=210$, $250$ and $282.5$~s respectively. Higher temperatures are reached inside the current sheet and plasmoids than during the laminar phase, as shown in panels E, L, S, G, N, and U from Fig.~\ref{fig:heating} by comparing with temperatures presented in Fig.~\ref{fig:heating_laminar}.

Additionally, higher temperatures are reached in MFMS than in TF, as illustrated in panels E, L, and S in comparison with panels G, N, and U of Fig.~\ref{fig:heating}. Indeed, in TF, the temperatures achieve a maximum of $35\,000$~K,  whereas in MFMS the temperatures can reach up to approximately $50\,000$~K. In TF, this temperature is reached with the H and H$^+$ drift heating (panels F, M, and T). These panels show that the rate of the heating terms ranges from $10^{-2.8}$ to $10^{-3}$~s$^{-1}$ inside the current sheet and plasmoids, increasing the temperature $T_{H^+}$ from $22\,000$~K (panel G) to $30\,000$~K. In the plasmoid located at $z\approx1.5$~Mm, the temperature achieves a maximum of $35\,000$~K due to the H-H$^+$ drift (panel U).

In MFMS, since many species and interactions are involved, the heating produced by drift velocities is more complex to interpret than TF and requires the analysis of several components of the heating terms during the plasmoid instability phase.

First, if we compare panels A, H, and O with F, M, and T, the rate associated with the heating terms due to the drift between H and H$^+$ is much higher in MFMS than in TF. The disparity can reach up to one order of magnitude in some regions of the current sheet. For example, at $z\approx1.6$ or $z\approx3.2$~Mm of panel O, the rate achieves a maximum at $10^{-2}$~s$^{-1}$. Therefore, the temperature $T_{H^+}$ increases in these regions from $30\,000$~K to $\sim50000$~K between $t=210$ and $t=282.5$~s, (panels E, L, and S). Even though most of the heating comes from the H and H$^+$ collision, the heating due to the H$^+$ and He$^+$ collision is not negligible at $z\approx1.6$ or $z\approx3.2$~Mm (panel R). This last heating rate is approximately $10^{-3.1}$~s$^{-1}$(panels D, K and R). 

Focusing on the formation of the plasmoid located at $z\approx1.9$~Mm in panels E, L, and S of Fig.~\ref{fig:heating}, one can notice that the thickness of this plasmoid increases. This dilation is the consequence of the balance of the thermal pressure of the plasma inside the current sheet that increases with respect to the magnetic pressure, which is constant. Therefore, during this dilation, the temperature inside the plasmoid is constant, and $\sim 50\,000$~
K, while its thickness increases from $\sim 0.08$~Mm to $\sim 0.2$~Mm. Therefore, the dilation process of the thermodynamic system defined as the volume of plasma located inside the plasmoid at $z\approx1.9$~Mm between $t=210$ and $t=282.5$~s is not adiabatic. However, if we look at panels A, H, and O or D, K, and R, it is clear that the heating terms due to the drift between H and H$^+$ or H$^+$ and He$^+$ are negligible inside the plasmoid. Therefore, one has to focus on other heating sources inside the plasmoids to explain the non-adiabaticity of the dilation.  

Indeed, the heating due to the friction between He and He$^+$ is not negligible inside the plasmoid, as shown in panels C, J, and Q of Fig.~\ref{fig:heating}. In panel J, at $z\approx 1.9$~Mm, the heating rate associated with He-He$^+$ collisions is approximately equal to $10^{-3.4}$~s$^{-1}$ which allows maintaining the plasma temperature inside the plasmoid to $50\,000$~K during the dilatation. The cooling rate associated with the adiabatic expansion of the plasmoids is negligible compared to this heating rate due to He-He$^+$ collisions ($\approx10^{-4.2}$~s$^{-1}$). The temperature of the second plasmoid located at $z\approx3$~Mm is slightly lower than the other plasmoid and ranges from $35\,000$~K to $40\,000$~K. This is because the initial plasma temperature in the current sheet where this plasmoid is initially formed is lower than the initial plasmoid temperature located at $z\approx 1.9$~Mm. Indeed, during the plasmoid instability phase, the temperature distribution is not uniform along the $z$ axis leading to some variations in the temperature of the formed plasmoids. 

Furthermore, the heating rate due to He and He$^+$ collisions in the plasmoid at $z\approx3$~Mm  is lower than at $z\approx 1.9$~Mm, and approximately equal to $10^{-4}$~s$^{-1}$, as we can see in the panel J of Fig.~\ref{fig:heating}. The heating due to the friction between He and H$^+$ slightly contributes to the heating of the plasmoids, as we can see in panels B, I, and P. Note that this contribution is not included in the TF model. Thus, during the expansion of the plasmoids, the temperature inside the plasmoids decreases and diffuses with time.

In summary, the temperature is higher in MFMS than in TF both during the laminar  (with temperatures of $30\,000$~K for MFMS and $20\,000$~K for TF) and plasmoid instability phases ($65\,000$~K for MFMS and $35\,000$~K for TF). Consequently, in MFMS and for the chosen initial thermodynamic conditions, the plasma reaches a sufficient temperature inside the current sheet to eventually ionize helium species rapidly. During the plasmoid instability phase, the main heating process is due to the collisions between H and H$^+$. It allows increasing the temperature inside the current sheet from $30\,000$~K to $50\,000$~K. Then, plasmoids are produced and expanded following a non-adiabatic process. In this context, neutral helium is rapidly ionized (as described with further details in Section~\ref{sec:fractionation}) and collides with all the other particles, in particular He, inside the plasmoids. It allows the maintenance of the temperature of the plasmoids at $50\,000$~K (for the plasmoid located at $z=1.8$~Mm). A maximum temperature of $65\,000$~K is achieved inside the plasmoids at a later time of the plasmoid instability phase.

\subsubsection{Decoupling of the temperatures in MFMS}
\label{sec:thermalnoneq}

In this section, we focus only on MFMS at a later time of the plasmoid instability phase. Indeed, these effects appear to be out of reach in the TF model. In MFMS, our previous results from Section~\ref{sec:heatingdrift} have shown that most of the temperatures are in equilibrium during the laminar and plasmoid instability phase. This is because the thermalization terms for any possible interactions $Q_{\specie\ions, \specie'\ions'}^{T,\text{col}},\,(\specie\ions,\specie'\ions')\in\mixture^2$ have larger magnitude than any other heating terms such as $Q_{\specie\ions, \specie'\ions'}^{\textbf{u},\text{col}},\,(\specie\ions,\specie'\ions')\in\mixture^2$ or $Q^{\text{spitz}}_{\elec}$. In other words, the characteristic timescale of the thermalization terms range from 10~s to just a few seconds whereas the heating terms associated with drift velocities range from 100~s to higher, as shown in Fig.~\ref{fig:heating}. 

Departures in temperature occur for a couple of species $(\specie\ions,\specie'\ions')\in\mixture^2$ if the rate associated with all the heating processes of particles $\specie\ions$ becomes much higher than the rate of the thermalization term $Q_{\specie\ions, \specie'\ions'}^{T,\text{col}}/e_{\specie\ions}$. In other words, for any $(\specie\ions, \specie'\ions')\in\mixture^2$, neglecting the electron thermal conduction and radiative losses, this condition can be written as

\begin{equation}
    \frac{Q_{\specie\ions, \specie'\ions'}^{T,\text{col}}}{e_{\specie\ions}} \ll \frac{1}{e_{\specie\ions}}\left(\sum_{\specie^*\ions^*\in\mixture} Q_{\specie\ions,\specie^*\ions^*}^{\textbf{u},\text{col}}\right)=\frac{Q_{\specie\ions}^{\textbf{u},\text{col}}}{e_{\specie\ions}},
    \label{eq:conditiontherm}
\end{equation}

\begin{figure*}[tbh]
	\begin{center}
	   \includegraphics[scale=0.32]{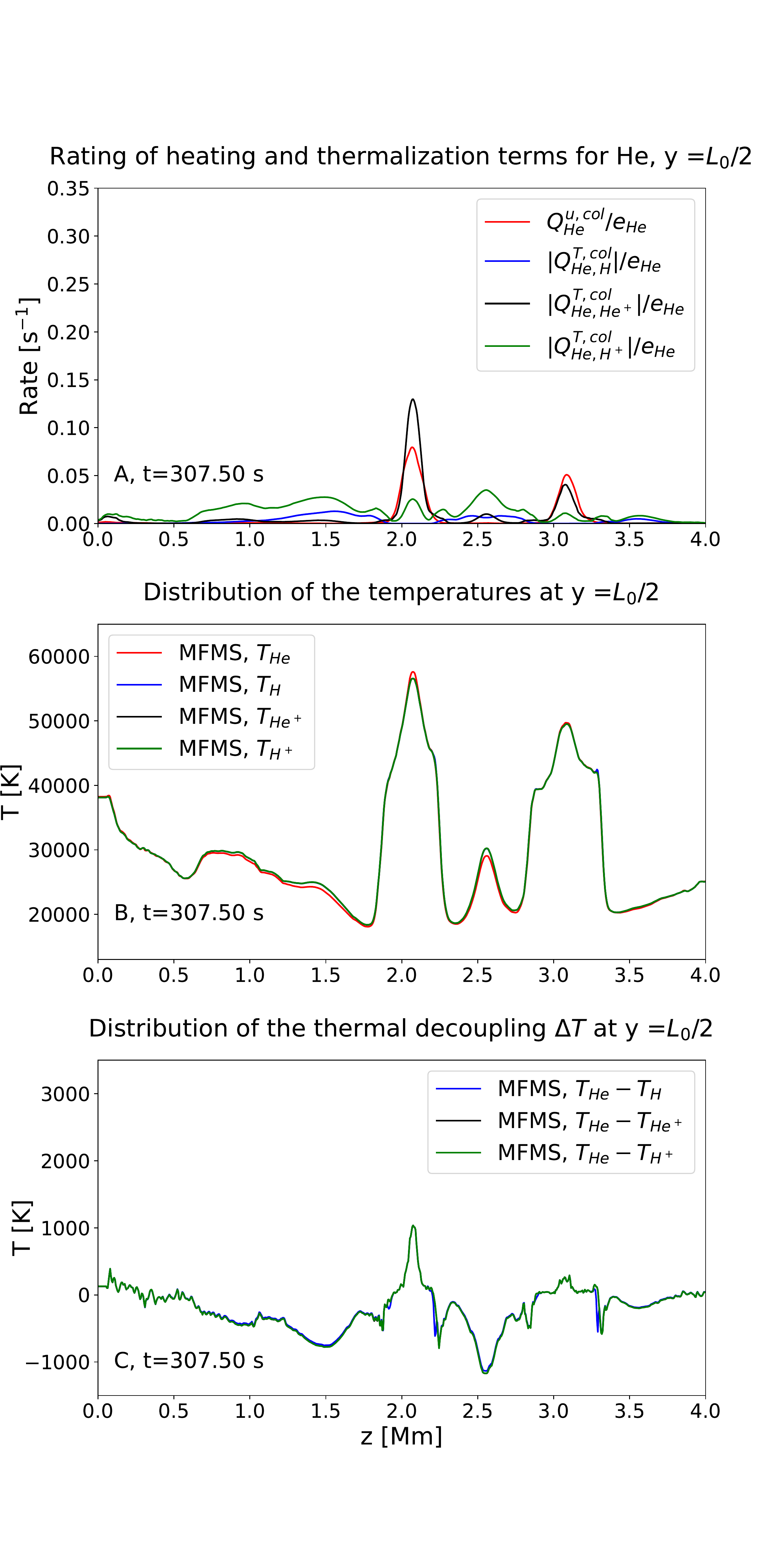}\vspace*{-0.2cm}
	\hspace*{-0.1cm}
	\includegraphics[scale=0.32]{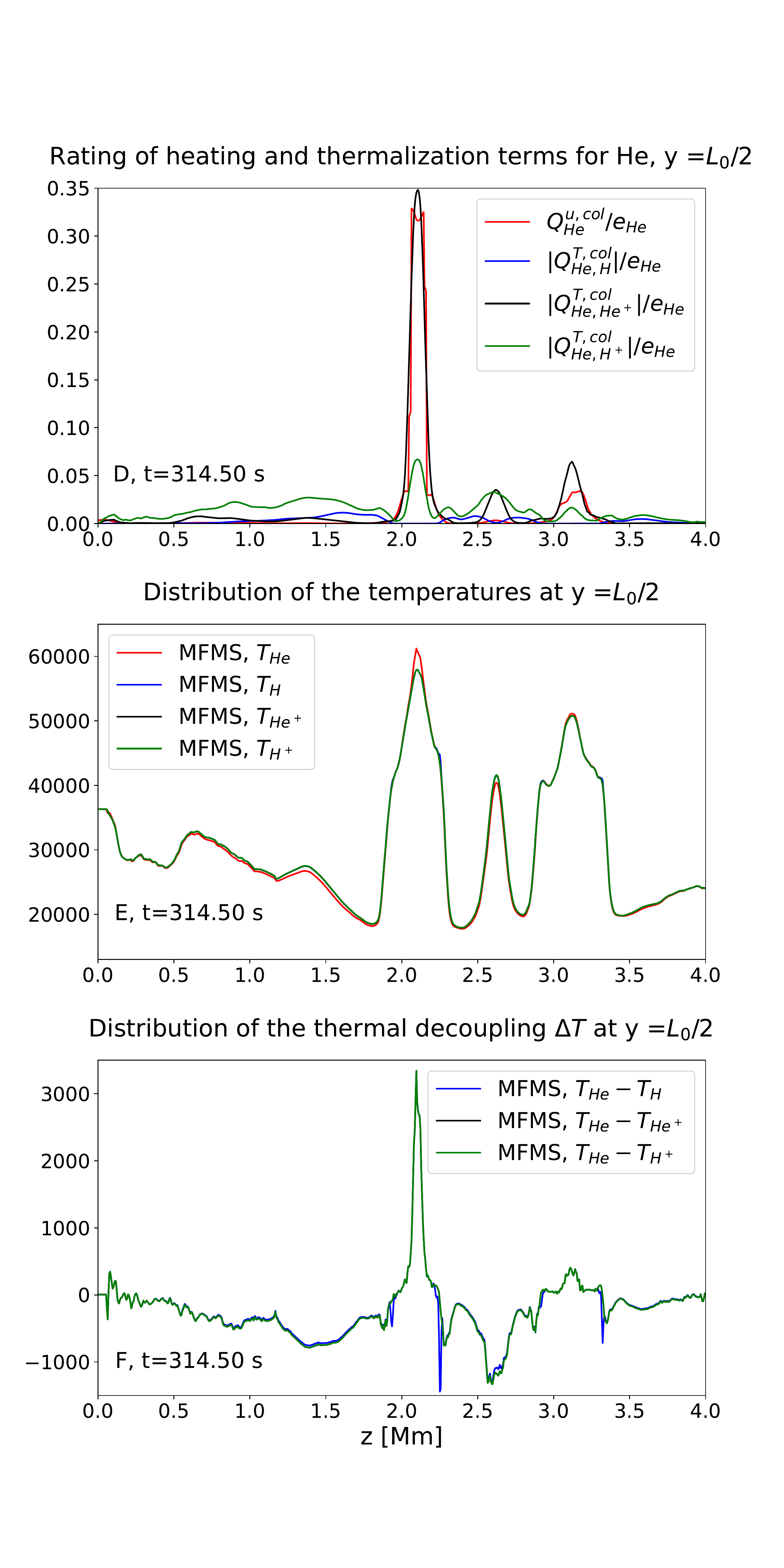}\vspace*{-1cm}
		\caption{From top to bottom row: distribution along the $z$ axis at $y=L_0/2$ of the rates of the total heating term $Q_{\text{He}}^{\textbf{u},\text{col}}/e_{\text{He}}$ and all the thermalization terms  $Q_{\text{He},\specie\ions}^{T,\text{col}}/e_{\text{He}}$, with $\specie\ions\in\mixture$, distribution of all the temperatures and distribution of temperature differences $\Delta T$ with respect to $T_{\text{He}}$ in the current sheet at position $y=L_0/2$ for MFMS. From left to right column: distributions at different times $t=307.5$~s and $t=314.5$~s.}
		\label{fig:thermnoneq}
	\end{center}
\end{figure*}

In order to highlight the thermal decoupling effects occurring during the reconnection process, we focus only on neutral helium. Neutral helium shows strong thermal decoupling effects due to the high decoupling in collisions with other species, as shown in the previous sections. Figure~\ref{fig:thermnoneq} shows the rate $Q_{\text{He}}^{\textbf{u},\text{col}}/e_{\text{He}}$ distribution compared with all the rates associated with the thermalization terms $Q_{\text{He}, \specie\ions}^{T,\text{col}}/e_{\text{He}},\, \specie\ions\in\mixture$ at times $t=307.5$~s, and $t=314.5$~s in the current sheet at $y=L_0/2$. On top of that, we also show the distributions of all the temperatures (panels B and E) as well as the differences in temperature $\Delta T=T_{\text{He}}-T_{\specie\ions},\,\specie\ions\in\heavy $ with respect to the neutral helium temperature $T_{\text{He}}$ (panels C and F).

First, if we focus on the plasmoid located at $z\approx2$~Mm at $t=307.5$~s, in panel A, the rate of the total heating term $Q_{\text{He}}^{\textbf{u},\text{col}}/e_{\text{He}}$ is slightly higher than the rates $Q_{\text{He}, \text{H}^+}^{T,\text{col}}/e_{\text{He}}$ and comparable to $Q_{\text{He}, \text{He}^+}^{T,\text{col}}/e_{\text{He}}$. After a few seconds, corresponding to the panel D of Fig.~\ref{fig:thermnoneq}, the rate of the total heating term $Q_{\text{He}}^{\textbf{u},\text{col}}/e_{\text{He}}$ increases drastically (500\% in 7 seconds) and becomes much larger than the rates of thermalization terms $Q_{\text{He}, \text{H}^+}^{T,\text{col}}/e_{\text{He}}$ and $Q_{\text{He}, \text{H}}^{T,\text{col}}/e_{\text{He}}$ and stay comparable to $Q_{\text{He}, \text{He}^+}^{T,\text{col}}$. In this framework, the condition in Eq.~\ref{eq:conditiontherm} is satisfied for the couples of species $(\text{He},\text{H}^+)$ and $(\text{He},\text{H})$. This leads to a fast and strong thermal decoupling where the departures in temperatures $T_{\text{He}}-T_{\text{H}^+}$ and $T_{\text{He}}-T_{\text{H}}$ achieve a maximum at $\approx3000$~K, occurring in just a few seconds, at the center of the plasmoid.

During the reconnection process, thermal decoupling effects can be stronger leading to a neutral helium temperature that can become much larger and reach up to $90\,000$~K where other species temperature stay at $70\,000$~K. Note that this effect occurs at later time of the plasmoid instability phase. Additionally, a small temperature decoupling is seen in neutral hydrogen, as illustrated with the blue line in panels C and F. However, this effect remains smaller than temperature decoupling effects associated with neutral helium.

\subsection{Evolution of the composition}
\label{sec:fractionation}

\subsubsection{Evolution of the ionization levels in MFMS and TF}
\label{sec:ionization}

Figures~\ref{fig:composition} (first two rows) and \ref{fig:composition_csheet} show the distribution of the total ionization fraction of MFMS and TF. Figure~\ref{fig:composition_csheet} is a cut in the middle of the current sheet at $y=L_0/2$ during the laminar phase (on the left) and plasmoid instability phase (on the right). In Fig.~\ref{fig:composition_csheet}, the distributions of the ionization fraction of hydrogen and helium species are shown. Figure~\ref{fig:ionrec} has the same layout as Fig.~\ref{fig:composition} but for the distribution of the ionization and recombination frequencies $\nu^{\text{ion}}_{\text{H},\text{H}^+}$, $\nu^{\text{rec}}_{\text{H}^+,\text{H}}$ for MFMS and TF, and $\nu^{\text{ion}}_{\text{He},\text{He}^+}$ for MFMS. These frequencies allows the assesment of the evolution of ionization and recombination effects and the composition of the plasma is altered during the reconnection process. The recombination frequency $\nu^{\text{rec}}_{\text{He}^+,\text{He}}$ is negligible and much smaller than the other ionization and recombination frequencies, thus, it is not shown here.

\begin{figure*}[tbh]
	\begin{center}
		\includegraphics[scale=0.35]{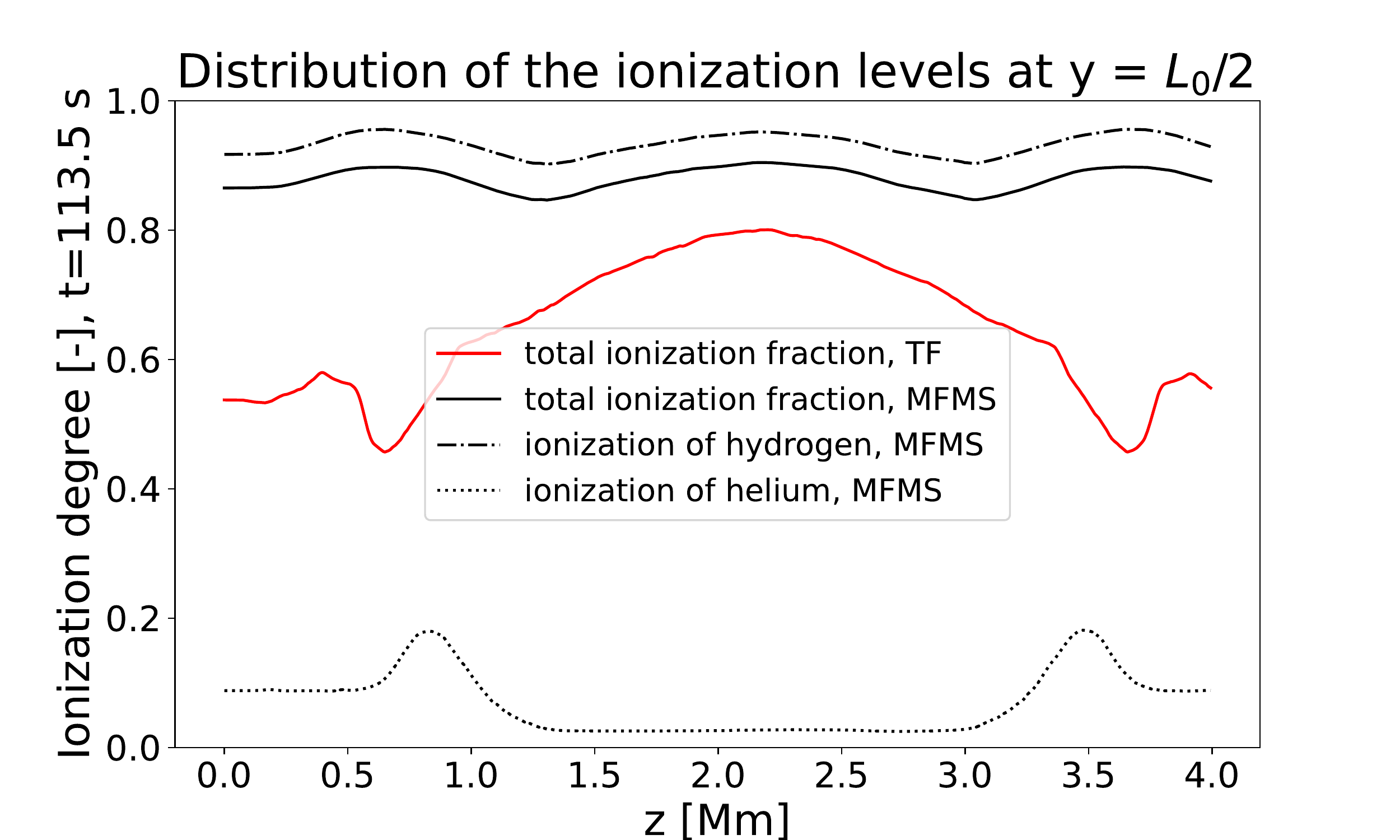}
		\includegraphics[scale=0.35]{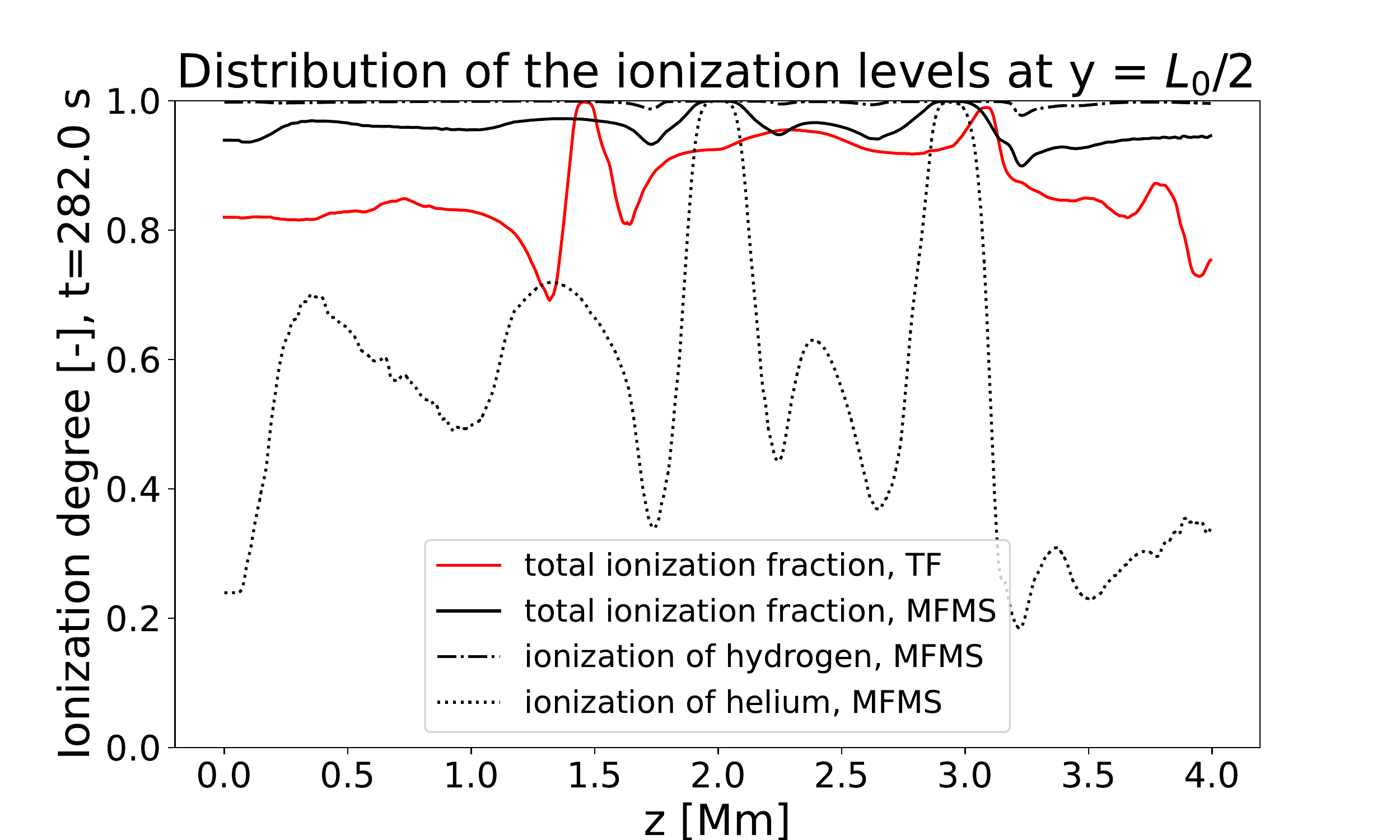}
		\caption{Distribution of the total ionization fraction for both MFMS and TF, ionization of hydrogen and helium species  in MFMS in the middle of the current sheet at position $y=L_0/2$ at times $t=113.5$~s (left) and $t=282$~s (right) respectively. The full black, dashed black and dot black lines corresponds to the total ionization fraction, the ionization of hydrogen species  $x_{\text{H}^+}=n_{\text{H}^+}/(n_{\text{H}^+} + n_{\text{H}})$ and the ionization of helium species $x_{\text{He}^+}=n_{\text{He}^+}/(n_{\text{He}^+} + n_{\text{He}})$ for MFMS. The full red line correspond to the total ionization fraction in TF. An animation of this figure is available.}
		\label{fig:composition_csheet}
	\end{center}
\end{figure*}

\begin{figure*}[tbh]
	\begin{center}
	 \hspace*{-1.7cm}
	\includegraphics[scale=0.4]{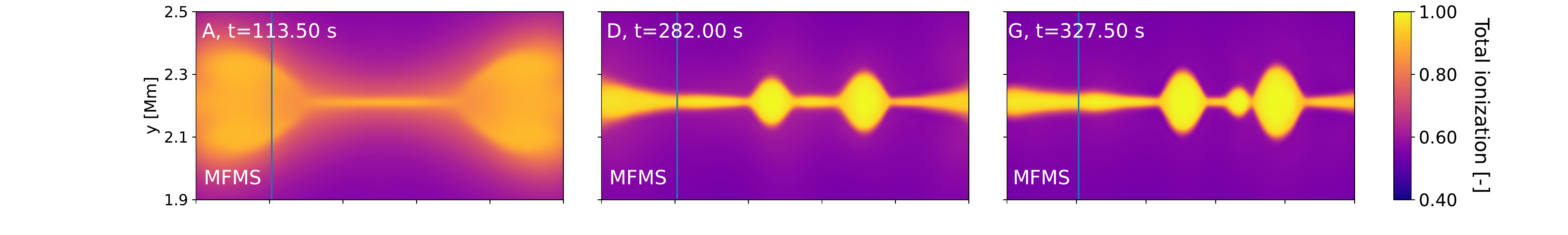}\vspace*{-0.2cm}\\
	\hspace*{-1.7cm}
	\includegraphics[scale=0.4]{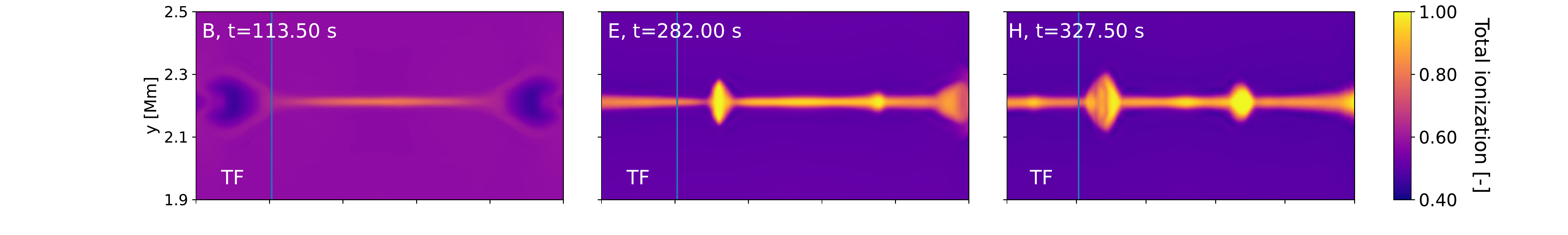}\vspace*{-0.2cm}\\
	\hspace*{-1.7cm}
	\includegraphics[scale=0.4]{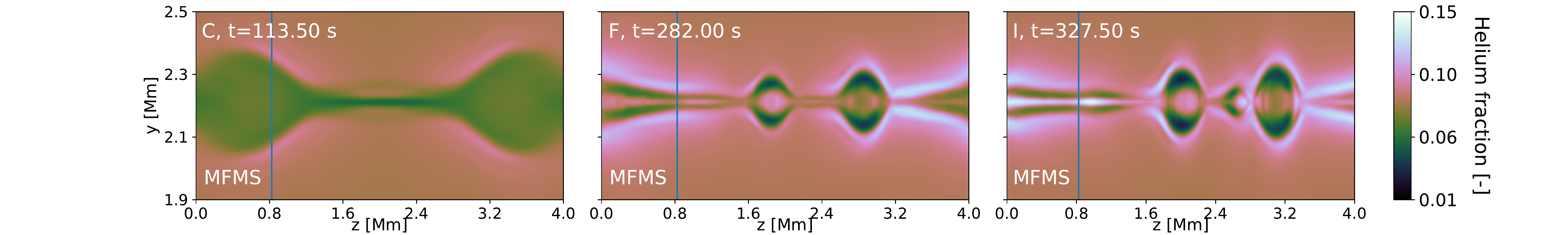}\vspace*{-0.2cm}
		\caption{Distribution of the total ionization fraction in TF (first row) and MFMS (second row), molar fraction of helium species (third row) $(n_{\text{He}}+n_{\text{He}^+})/(\sum_{\specie\ions\in\mixture}n_{\specie\ions})$ in MFMS, at different times $t=113.5$~s, \,$t=282$~s and $327.5$~s for any $(y,z)\in[1.8,2.6]\times[0,L_0]$~Mm (from left to right column) respectively. The vertical line corresponds to $z=0.825$~Mm. An animation of this figure is available.}
		\label{fig:composition}
	\end{center}
\end{figure*}

 \begin{figure*}[tbh]
	\begin{center}
\hspace*{-1.7cm}
	\includegraphics[scale=0.4]{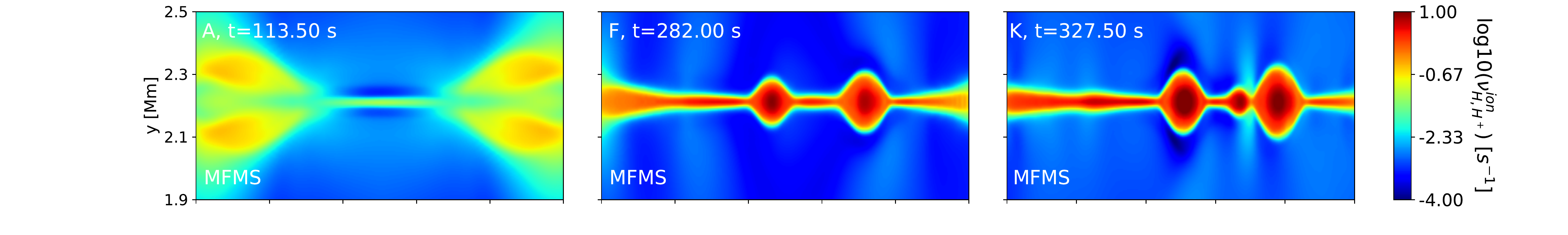}\vspace*{-0.2cm}\\
	\hspace*{-1.7cm}
	\includegraphics[scale=0.4]{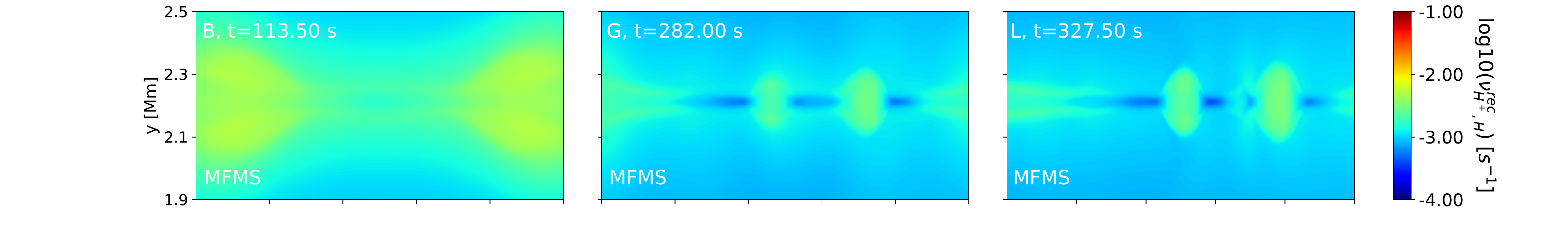}\vspace*{-0.2cm}\\
	\hspace*{-1.7cm}
	\includegraphics[scale=0.4]{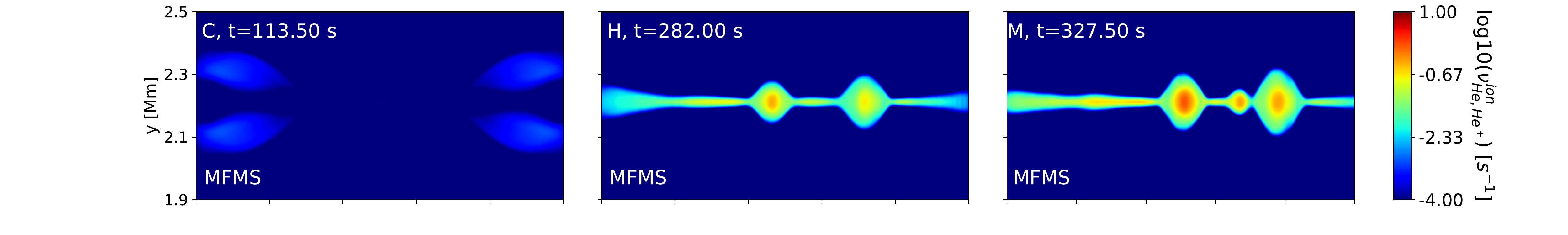}\vspace*{-0.2cm}\\
	\hspace*{-1.7cm}
	\includegraphics[scale=0.4]{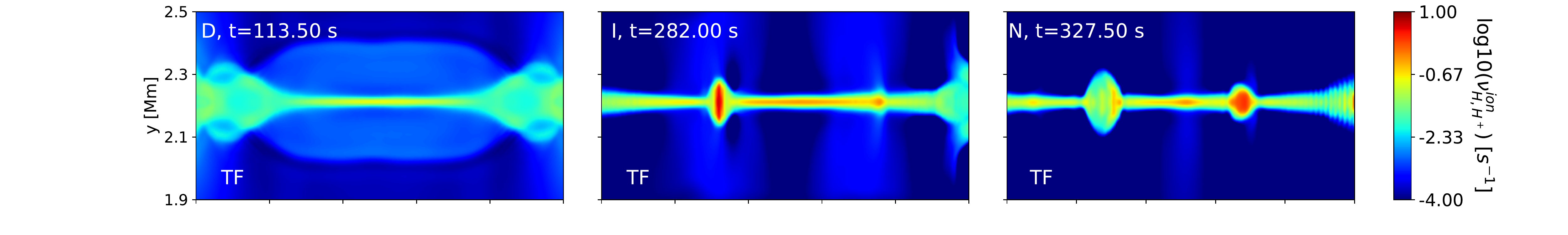}\vspace*{-0.2cm}\\
	\hspace*{-1.7cm}
	\includegraphics[scale=0.4]{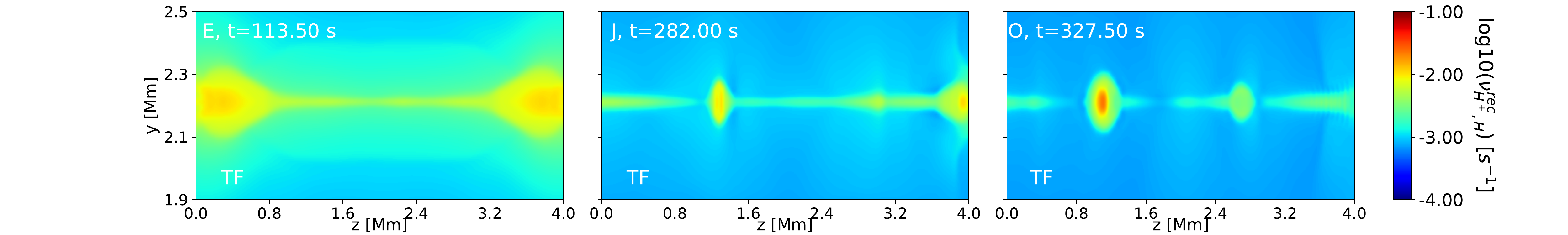}\vspace*{-0.2cm}
		\caption{Distribution of the decimal logarithm of the ionization and recombination frequencies in s$^{-1}$ of MFMS (first three rows) and TF (last two rows)
		at different times $t=113.5$~s, \,$t=282$~s and $327.5$~s for any $(y,z)\in[1.8,2.6]\times[0.4,4]$~Mm (from left to right column) respectively. From top to bottom row: distribution of $\nu^{\text{ion}}_{\text{H},\text{H}^+}$, $\nu^{\text{rec}}_{\text{H}^+,\text{H}}$ and $\nu^{\text{ion}}_{\text{He},\text{He}^+}$ in MFMS, and distribution of $\nu^{\text{ion}}_{\text{H},\text{H}^+}$, $\nu^{\text{rec}}_{\text{H}^+,\text{H}}$ in TF. Note that the distribution of  $\nu^{\text{rec}}_{\text{He}^+,\text{He}}$ is not shown since the magnitude is negligible and much smaller than $10^{-4}$~s$^{-1}$. An animation of this figure is available.}
		\label{fig:ionrec}
	\end{center}
\end{figure*}

During the laminar phase, represented in the first column of Figures~\ref{fig:composition} (panels A and B) and \ref{fig:ionrec} (panels A to E), and left plot of Fig.~\ref{fig:composition_csheet}, the ionization level is increasing in the current sheet for both MFMS and TF due to the ionization of hydrogen species. Indeed, the ionization frequency $\nu^{\text{ion}}_{\text{H},\text{H}^+}$ in the current sheet, in both MFMS and TF, is higher by more than one order of magnitude than the recombination frequency $\nu^{\text{rec}}_{\text{H}^+,\text{H}}$, as shown in Fig.~\ref{fig:ionrec}. Furthermore, the recombination frequency $\nu^{\text{rec}}_{\text{H}^+,\text{H}}$ in MFMS (panel B) is lower than in TF (panel E). This leads to an ionization level that is much higher in MFMS (ranging from $85\%$ to $95\%$) than in TF (ranging from $70\%$ to $80\%$) during the laminar phase, as shown on the left of Fig.~\ref{fig:composition_csheet}. These disparities obtained in the ionization and recombination frequencies between MFMS and TF are attributed to the difference in temperature of the plasma in the current sheet during the laminar phase. Indeed, as described in the previous Section~\ref{sec:heating}, the presence of helium species that collide with hydrogen species in MFMS leads to higher temperature in the current sheet than in TF. This increase of temperature gives rise to a larger hydrogen ionization in MFMS than in TF during the laminar phase.

The ionization levels in plasmoids are drastically increasing for both MFMS (right panel in Fig.~\ref{fig:composition_csheet}) and TF compared to the laminar phase (left panel). In TF, we can distinguish three plasmoids located at $y=1.5$~Mm,\,3.1~Mm and 3.7~Mm, and, in MFMS, two plasmoids located at $y=1.9$~Mm and 3~Mm. In TF, at the middle of the plasmoid located at $y=1.5$~Mm, the ionization level increases up to more than 98\%. This is attributed to a local increase of the temperature inside plasmoids, giving rise to an increase of H$^+$ due to ionization process, as illustrated in panel I of Fig.~\ref{fig:ionrec}.

In MFMS, the friction of hydrogen species with neutral helium allows a higher temperature level in the plasmoids than in TF during the plasmoid instability phase, as described in the previous Section~\ref{sec:heating}. This level of temperature is high enough to ionize helium, as illustrated in Fig.~\ref{fig:composition_csheet} characterized by a large increase of the black dot line. This is also illustrated in panels H and M of Fig.~\ref{fig:ionrec} where we can clearly see that $\nu^{\text{ion}}_{\text{He}^+,\text{He}}$ increases drastically, especially in the core of plasmoids. At this moment, a self-sustaining mechanism occurs. Indeed, when He$^+$ species are produced, they collide with both H$^+$ and He, as described in Section~\ref{sec:heating}. In this context, an additional heating process occurs due to collisions involving He$^+$. This effect will slightly increase the temperature at the center of the plasmoid, leading to even more He$^+$ due to the ionization process. This is characterized by the increase of $\nu^{\text{ion}}_{\text{He}^+,\text{He}}$ in the core of plasmoids, as illustrated in panels H and M of Fig.~\ref{fig:ionrec}.

\subsubsection{Chemical fractionation in MFMS}

In this section, we focus only on MFMS and chemical fractionation between helium and hydrogen species during the reconnection process. In Fig.~\ref{fig:composition} (last row), the distributions of the helium species molar fraction at different times, corresponding to the laminar and plasmoid instability phase, is shown. The molar fraction is defined as  $(n_{\text{He}}+n_{\text{He}^+})/(\sum_{\specie\ions\in\mixture}n_{\specie\ions})$.

During the laminar phase, in MFMS, the helium species molar fraction (panel C of Fig.~\ref{fig:composition}) decreases in the current sheet (from 7.9\% to 6\%). Additionally, we also notice a small increase of the helium species molar fraction at the boundary of the current sheet close to the outflow regions. Indeed, since helium is not ionized, these effects are attributed to the accumulation of neutral helium species which are decoupled from hydrogen species nearby the current sheet, as described in Section~\ref{sec:decoupling}. 

During the plasmoid instability phase, the helium species molar fraction increases (from 6\% to 10\%) in the core of plasmoids, as shown in panel F of Fig.~\ref{fig:composition}). This separation process, also called chemical fractionation, occurs inside the plasmoids and remains even when the latter are advected along the $z$ axis. This separation process occurs due to the ionization of helium species during the instability phase.
 
At a later time of the plasmoid instability phase, the chemical fractionation process gets stronger, leading to more helium species inside the current sheet that are advected towards outflow regions. This effect is illustrated in panel I of Fig.~\ref{fig:composition}. Indeed, an increase of helium species, up to 12\%, in the middle of the current sheet at $y=L_0/2$ and $z\in[0.4,1.5]$~Mm is shown. Note that the fusion of plasmoids also leads to a large increase of helium abundance as we can see in the same figure, in the current sheet at $z=2.9$~Mm.

To characterize the chemical fractionation in the outflows, the fluxes in number densities of helium and hydrogen species have been calculated and compared at the exhaust of the reconnection event until $t_{\text{end}}=454$~s. These calculations have been performed at a fixed $z=0.825~$~Mm. For any species $\specie\ions\in\mixture$, its corresponding flux in number density is calculated as

\begin{equation}
    \phi\left(\specie\ions\right) = dt\, dx \int_{l_y}\left[u_{z,\specie\ions}n_{\specie\ions}\right]\left(y,\,z=0.825~Mm\right) dy,
\end{equation}
where $dx$ is the grid size in the $x$ direction and set to one, $l_y$ is the length in the $y$ direction limited to the outflow region. We define the outflow regions as the FWHM of the maximum outflow at $z=0.825~$~Mm. Similarly as the thickness of the current sheet in Section~\ref{sec:currentsheet}, this length has been calculated by locating the maximum outflow velocity at $z=0.825~$~Mm. Then, we consider the termination of the exhaust at approximately $90\%$ of this maximum in the $y$ direction to estimate $l_y$.

Figure~\ref{fig:fluxes} shows the time evolution of the maximum and mean total outflow velocity (left) and the ratio of flux of helium species with respect to the flux of all species $\phi(\text{He}+\text{He}^+)/(\phi(\text{He}+\text{He}^+)+\phi(\text{H}+\text{H}^+))$ (right). In Fig.~\ref{fig:fluxes}, if the ratio is above 0.079 (corresponding to the initial total number density of helium species 7.9\% shown in Table~\ref{tab:table}) means helium enrichment and, otherwise, hydrogen enrichment. 

Our calculations show that during most of the reconnection process, the flux in the number density of helium species represents from $6$ to $8\%$ of the total flux of all species, as shown in the bottom left panel of Fig.~\ref{fig:fluxes}. Helium is enhanced at the exhaust when the helium ionization starts at $t\approx 250$~s during the plasmoid instability regime. It increases the helium flux up to 14\% of the total flux, as shown in the bottom left panel of Fig.~\ref{fig:fluxes}. These calculations are consistent with our figures presented in Fig.~\ref{fig:composition}. The helium enrichment in the outflows during the strongest flows and late-formed plasmoids also depends on the amount of helium species collected during the formation of plasmoids and the current sheet. In some localized events on the right-hand side of the reconnection, we found the opposite behavior when He has been previously deployed, and hydrogen is left in the latest plasmoids leading to the opposite, \ie\, enrichment of hydrogen.

Note that, from $t\approx 250$~s to $t_{\text{end}}=454$~s, strong variations of the maximum and mean total outflow velocity at the exhaust appear, as shown in the top left of Fig.~\ref{fig:fluxes} ranging between 15~km s$^{-1}$ to 24~km s$^{-1}$ for the maximum outflow velocity. These peaks coincide with peaks in the flux of helium species with respect to the total flux, as shown in the bottom left of Fig.~\ref{fig:fluxes}. We note that helium species are mostly ionized at the exhaust, as illustrated in Fig.~\ref{fig:composition_csheet} on the right.

\begin{figure*}[tbh]
	\begin{center}
	\hspace*{-0.4cm}
		\includegraphics[scale=0.35]{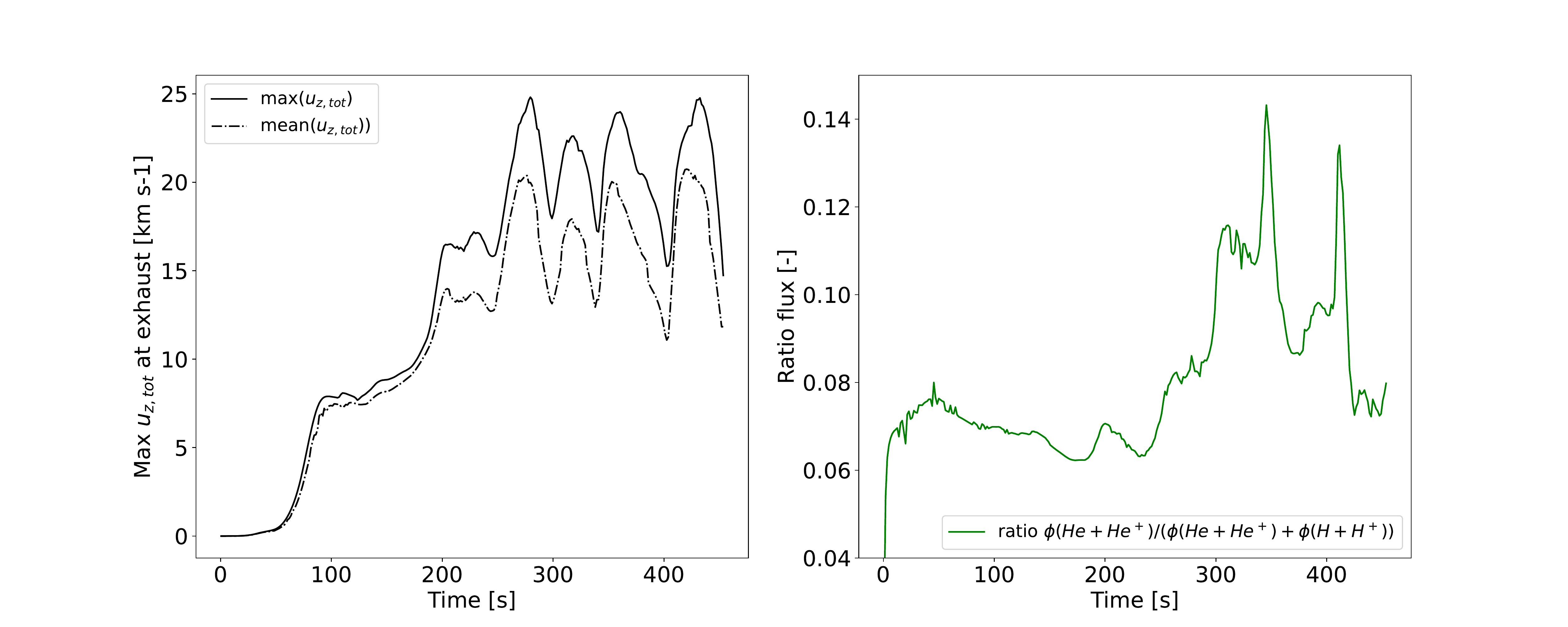}
		\vspace*{-0.3cm}
		\caption{Distribution of the evolution in time of the maximum and mean total outflow velocity (left) and the ratio of fluxes $\phi(\text{He}+\text{He}^+)/(\phi(\text{He}+\text{He}^+)+\phi(\text{H}+\text{H}^+))$ (right) for MFMS at the exhaust. On the left figure, the full and dashed black lines correspond to the maximum and mean total outflow velocity.}
		\label{fig:fluxes}
	\end{center}
\end{figure*}

Taking into account the complexity of the chemical evolution, chromospheric reconnection may play a role in helium enrichment. In short, at the laminar phase, the ionization fraction for hydrogen is higher, and it is slowly expelled. At that time, outflows are relatively small. However, during the second phase, with stronger flows and plasmoids, the number of helium atoms is larger and helium is ionized due to the temperature increase, resulting in an enrichment of helium at the outflows. However, as mentioned above, in some locations of the right-hand side of the reconnection, stronger flows also could produce hydrogen enrichment depending on the history of the plasma evolution inside the current sheet, so care must be taken when interpreting these results.

\section{Conclusions and Discussion}~\label{sec:con}

In order to study a magnetic reconnection event under upper chromospheric conditions, we have focused on a multi-fluid multi-species (MFMS) MHD model and compared it with a two-fluid (TF) MHD model using \ebysus\ code. The MFMS model is an extension of the TF approach, generalized to any possible species with different ionization levels, considered as separate fluids and coupled altogether with  collisions and ionization or recombination processes. Both models include anisotropic electron thermal conductivity and radiative losses. In this work, we have focused on two different mixtures: a helium-hydrogen mixture (MFMS model) $\mixture$ and a hydrogen or two-fluid (TF model) mixture $\mixture_{\text{TF}}$. By doing so, our strategy allows the comparison of the classical two-fluids approach \citep[see][]{Leake:2012pr,Leake:2013qz,Murtas2021} with a MFMS approach and highlight the role of the helium species in a chromospheric reconnection event.  Unlike previous studies of chromospheric magnetic reconnection from the literature \citep[see][]{Leake:2012pr,Leake:2013qz,Murtas2021}, the collisional frequencies are not assumed constant but are calculated accurately based on a formalism with collision integrals consistent at the kinetic level with the 13N moment model, as described with further details in \citet{Wargnier2022}. 


For the two approaches TF and MFMS, we have considered a perturbed Harris current sheet in upper chromospheric conditions, inspired by \citet{Leake:2012pr,Leake:2013dq}. We have compared and analyzed results in both cases at different levels: the evolution of the total current sheet, the velocity fields and decoupling of the particles, the composition and evolution of the ionization levels in MFMS and TF, and the main heating processes.

For both cases, we have two different phases: 1- a laminar phase where an elongated current sheet is formed and 2- a plasmoid instability phase where many plasmoids are formed  and are advected from the middle of the reconnection region towards the outflow regions. Our calculations have shown that the evolution of the current sheet and the structure of the total velocity fields slightly differs between MFMS and TF. Indeed, in MFMS, the presence of helium species slightly slows the flow, leading to slightly smaller velocities in MFMS than in TF. Additionally, the average thickness of the current sheet and plasmoids is larger in MFMS than in TF while the average length of the current sheet is similar between the two.

During the laminar phase, outflow and inflow velocities are formed due to the Lorentz force generated by the initial structure of the magnetic field. In both MFMS and TF, velocities reach $20$~km s$^{-1}$ and $3.5$~km s$^{-1}$ in the $z$ and $y$ directions, respectively. In both directions, in MFMS, our calculations have shown that the drift velocity between H$^+$ and He is much higher than any other drift velocities. This decoupling is due to the fact that neutrals He decouples from H and H$^+$ when the thickness of the current sheet thins down to the order of the mean free path between He and H$^+$ or H, and, neutral He atoms are not dragged into the current sheet as quickly as H and H$^+$. The fact that He decouples from H and H$^+$ and not H with H$^+$ is attributed to smaller collisional rate for He-H$^+$ or He-H interactions than for H-H$^+$ interactions in upper chromospheric conditions. We refer to \citet{Wargnier2022} for further details about the magnitude of these collisional rates. The relatively large He and H$^+$ drift velocity in MFMS does not occur in the TF model since the latter considers only hydrogen species, \ie\, H and H$^+$.

During the plasmoid instability phase, the amplitude of all the drift velocities is increasing in both MFMS and TF. The amplitude of the drift velocity associated with He-H$^+$ interactions is the highest inside plasmoids. In MFMS, the drift between H and H$^+$ is stronger at the boundary of plasmoids and inside the current sheet, while it is approximately three times smaller and shows an oscillating pattern parallel to the current sheet in TF. In both MFMS and TF, the decoupling between H and H$^+$ occurs because the thickness of the current sheet continues to thin down to the order of the mean free path between H and H$^+$, and, neutral H does not collide enough with H$^+$ during the plasmoid instability phase, similarly to He-H$^+$ interactions in MFMS during the laminar phase. The amplitude of the drift between H and H$^+$ is smaller in TF than in MFMS - meaning that the collisional coupling is stronger in TF than in MFMS due to the presence of helium species, as described with further details in  Section~\ref{sec:decoupling}.

In both models, the main heating processes are due to the drift velocities between any possible particles, also written as $Q_{\specie\ions,\specie'\ions'}^{\textbf{u},\text{col}}$. During the laminar phase, our calculations have shown that adding helium in the model allows to considerably increase the temperature of the plasma in the current sheet compared to TF. Inside the current sheet, the temperature is approximately 28000~K in MFMS and 19000~K in TF. In this context, a sufficient level of temperature is reached due to collisions associated with H-H$^+$ in the current sheet to ionize helium. Finally, collisions associated with He-He$^+$ and He-H$^+$ produce a small additional heating in the core of plasmoids and allows to maintain the temperature level to 60000~K whereas only 30000~K is reached with TF. Note that since the electrons are strongly coupled by collisions with H$^+$ species, the maximum electron temperature reached is also at 60000~K.

In the MFMS approach, due to the weak coupling between neutral helium and other species, strong thermal decoupling effects have been obtained at a later time of the plasmoid instability phase. Our results have demonstrated that the temperature of neutral helium inside plasmoids can increase drastically up to $90\,000$~K while the temperatures of other species reach a maximum of $70\,000$~K. Note that a potential improvement of these models would be to include photoionization of helium species, as highlighted by \citet{Golding_2014, Golding_2016}.

By combining a model where the collisional rates have been calculated accurately with an efficient numerical strategy, we have demonstrated that in upper chromospheric magnetic reconnection the inclusion of helium species in the system allows a considerable increase temperature of the plasma while the flows are weaker than a model where only hydrogen species are considered. This large increase of temperature compared to a classical two-fluid case is associated with a weak coupling between neutral helium and hydrogen that leads to a high energy release due to large drift velocities in the current sheet and inside plasmoids. In addition, during the plasmoid instability phase, inside the plasmoids, two efficient mechanisms can occur and allow an additional increase of the temperature: the production of He$^+$ due to ionization and recombination processes and collisions with other species, and thermal decoupling effects which can increase the temperature of neutral helium species locally. These mechanisms are able to easily increase the temperature of the plasma during the plasmoid instability regime of a magnetic reconnection event from upper chromospheric temperatures ($16\,000$~K) to transition region temperatures ($90\,000$~K) on typical spatial and temporal scales of Megameters and minutes respectively.

During the laminar phase, the total ionization level is much higher in MFMS (from 85\% to 95\%) than in TF (from 70\%) to 80\%) due to more efficient heating mechanisms in MFMS, as described previously, leading to higher temperature in the current sheet giving rise to a large increase of the total ionization level, mostly populated by H$^+$. During the plasmoid instability phase, in both MFMS and TF, the total ionization level increases drastically in the current sheet and plasmoids. In MFMS, both neutral hydrogen and helium species are ionizing, thus, the plasmoids and current sheet are mostly populated by H$^+$ and He$^+$. In TF, the total ionization level inside the plasmoids, mostly due to the presence of H$^+$ species, is slightly smaller than in MFMS and is ranging from 80 to 98\%. 


Note at a later time of the plasmoid instability phase, chemical fractionation effects have been captured in MFMS. Indeed, as shown in Section~\ref{sec:fractionation}, helium and hydrogen species are separated inside the current sheet, characterized by an increase of helium species (from 5 to 12\%) and a decrease of hydrogen species (from 98\% to 89\%), which are advected in the outflow directions, along the $z$ axis. During the plasmoid instability phase, the temperature increase inside the current sheet and results in the ionization of helium species leading to helium species enrichment at the exhaust of the reconnection.



Our study suggest that reconnection could lead to an enrichment of helium during the strong outflow regimes under certain conditions. This mechanism, out of reach of single-fluid or two-fluid MHD approaches, might explain the enhancement of helium species observed at upper layers of the solar atmosphere associated to the switchbacks \citep{Fargette_2021,Bale_2021} from the upper chromosphere. Similarly, intriguing observations of enhanced Helium abundance in CMEs have been linked to reconnection events in the chromosphere \citep{Fu2020}. Our results suggest that in the upper layer of the solar atmosphere, any event associated with a large variation or reconfiguration of the magnetic field could in principle release a sufficient level of energy to produce this chemical fractionation between hydrogen and helium species while increasing the temperature of the plasma.
However, a simulation of a full stratified atmosphere with a helium-hydrogen mixture, including alpha particles, while guaranteeing that most of the mean free paths in that mixture are spatially resolved, would be necessary to further investigate this issue.


\section{Acknowledgments}

We gratefully acknowledge support by NASA contracts, NNX17AD33G, 80NSSC18K1285, 80NSSC20K1272, and NNG09FA40C (IRIS), NSF contract AST1714955. Resources supporting this work were provided by the NASA High-End Computing (HEC) Program through the NASA Advanced Supercomputing (NAS) Division at Ames Research Center. The simulations have been run on clusters from the Notur project, and the Pleiades cluster through the computing project s1061, s2053 and s8305. We thankfully acknowledge the support of the Research Council of Norway through contract 230938/F50 and through grants of computing time from the Programme for Supercomputing.

\appendix
\section{Definition of the MFMS resistivity in small drift velocities asymptotic limit}
\label{sec:resistivity}
By considering the definition of the electric field and electron velocity from Eqs.~\ref{eq:e_ohm} and \ref{eq:vele} it is possible to provide a new definition of the resistivity of the MFMS model in terms of collision integrals $\Omega^{1,1}_{\elec,\specie\ions}(\mu_{\elec,\specie\ions}),\, \specie\ions\in\heavy$ following \citet{Wargnier2022}. Indeed, assuming that the mass of the electrons is much smaller than the mass of any heavy particles, the collisional frequencies between $\elec$ and any $\specie\ions\in\heavy$ reads

\begin{equation}
    \nu^{\text{col}}_{\elec,\specie\ions} = \tfrac{4}{3}\tfrac{m_{\elec,\specie\ions}}{m_{\elec}} n_{\specie\ions} |\thermalspeed_{\elec,\specie\ions}|\, \Omega^{1,1}_{\elec,\specie'\ions'}(\mu_{\elec,\specie\ions}) \approx \tfrac{4}{3} n_{\specie\ions} |\thermalspeed_{\elec}|\, \Omega^{1,1}_{\elec,\specie'\ions'}(\mu_{\elec}),
    \label{eq:simplify}
\end{equation}
where the thermal speed is simply defined as $|\thermalspeed_{\elec}| = \sqrt{8/\pi\mu_{\elec}}$. If we consider the definition of Eq.~\ref{eq:simplify} and the third term of Eq.~\ref{eq:e_ohm} we obtain

\begin{equation}
  -\frac{1}{n_{\elec} q_{\elec}} \momex^{\text{col}}_{\elec} = \tfrac{4}{3} \tfrac{m_{\elec}|\thermalspeed_{\elec}|}{q^2_{\elec}\, x_{\ionized}} \sum_{\specie\ions\in\heavy} x_{\specie\ions} \Omega^{1,1}_{\elec,\specie\ions}(\mu_{\elec})\left[\sum_{\specie'\ions'\in\heavy}n_{\specie'\ions'}q_{\specie'\ions'}\velocity_{\specie'\ions'} -\totalcurrent -n_{\elec}q_{\elec}\velocity_{\specie\ions}\right].
  \label{eq:resistiveterm}
\end{equation}
where $x_{\ionized} = n_{\elec}/(n_{\elec} + \sum_{\specie\ions\in\heavy}n_{\specie\ions})$ is the ionization fraction of the whole plasma and $x_{\specie\ions}=n_{\specie\ions}/(n_{\elec} + \sum_{\specie'\ions'\in\heavy}n_{\specie'\ions'})$ is the molar fraction of particle $\specie\ions\in\heavy$. Eq.~\ref{eq:resistiveterm} can be interpreted as the resistive term of the MFMS model. In the limit where drift velocities are smaller than the hydrodynamic velocity of the plasma or if the plasma is fully ionized or if the plasma is composed only of one type of ionized species Eq.~\ref{eq:resistiveterm} simplifies to 
\begin{equation}
\frac{1}{n_{\elec} q_{\elec}} \momex^{\text{col}}_{\elec} \approx \left[\tfrac{4}{3} \tfrac{m_{\elec}|\thermalspeed_{\elec}|}{q^2_{\elec}\, x_{\ionized}} \sum_{\specie\ions\in\heavy} x_{\specie\ions} \Omega^{1,1}_{\elec,\specie\ions}(\mu_{\elec})\right] \totalcurrent = \eta \totalcurrent,
\end{equation}
where the resistivity is defined as $\eta = \tfrac{4}{3} \tfrac{m_{\elec}|\thermalspeed_{\elec}|}{q^2_{\elec}\, x_{\ionized}} \sum_{\specie\ions\in\heavy} x_{\specie\ions} \Omega^{1,1}_{\elec,\specie\ions}(\mu_{\elec})$. Therefore, we have been able to define the resistivity of the MFMS model in terms of collision integrals for any possible mixture valid in the asymptotic limit where where the drift velocities are assumed to be much smaller than the hydrodynamic velocities of the MFMS plasma.  Note that  this resistivity coefficient is always true for the TF case based on $\mixture_2{\text{TF}}$.

\section{Compact form of the MFMS governing equations}
\label{app:system}

The presented system \ref{eq:continuity},\ref{eq:spmoma1},\ref{eq:enei5},\ref{eq:enee5} and \ref{eq:magne} of the MFMS model can be rewritten in a compact form. By doing so, we can easily distinguish the nature of the different terms involved in the system. The system can be rewritten as 
\begin{equation}
\dt \mathrm{U} +  \nabla \cdot \mathrm{F}\left(\mathrm{U}\right) +  \nabla \cdot \mathrm{D}\left(\mathrm{U},\nabla \mathrm{U}\right) = \mathrm{S}^{\text{conv}}\left(\mathrm{U},\nabla \mathrm{U}\right)+\mathrm{S}^{\text{stiff}}\left(\mathrm{U}\right),
\label{eq:fullsystem}
\end{equation}
\noindent where $\mathrm{U}\in \mathbb{R}^{5 \mathfrak{n}_{\mathrm{\heavy}}+4}$, $\mathrm{F}\left(\mathrm{U}\right)\in\mathbb{R}^{3\times(5 \mathfrak{n}_{\mathrm{\heavy}}+4)}$, $\mathrm{D}\left(\mathrm{U},\nabla \mathrm{U}\right)\in\mathbb{R}^{3\times(5 \mathfrak{n}_{\mathrm{\heavy}}+4)}$, 
$\mathrm{S}^{\text{conv}}\left(\mathrm{U},\nabla \mathrm{U}\right)\in\mathbb{R}^{5 \mathfrak{n}_{\mathrm{\heavy}}+4}$ and $\mathrm{S}^{\text{stiff}}\left(\mathrm{U}\right)\in\mathbb{R}^{5 \mathfrak{n}_{\mathrm{\heavy}}+4}$. These terms read
\begin{eqnarray}
    \mathrm{U} &=& \left((\rho_{\specie\ions})_{\specie\ions\in\heavy},\;(\rho_{\specie\ions} \velocity_{\specie\ions}^T)_{{\specie\ions}\in\heavy},\; (e_{\specie\ions})_{{\specie\ions}\in\heavy},\; e_{\elec},\;\BField^T \right)^T, \nonumber\\
    \mathrm{F}\left(\mathrm{U}\right) &=& \left((\rho_{\specie\ions} \velocity_{\specie\ions})_{{\specie\ions}\in\heavy},\;(\rho_{\specie\ions} \velocity_{\specie\ions}\otimes\velocity_{\specie\ions} + P_{\specie\ions}\identity)_{{\specie\ions}\in\heavy},\; (e_{\specie\ions}\velocity_{\specie\ions})_{{\specie\ions}\in\heavy},\; e_{\elec}\velocity_{\elec},\;\identity\pvec\left(\velocity_{\elec}\pvec\BField\right)\right)^T, \nonumber\\
    \mathrm{D}\left(\mathrm{U},\nabla \mathrm{U}\right) &=& \left(0_{\mathfrak{n}_{\mathrm{\heavy}}},\; 0_{3\times\mathfrak{n}_{\mathrm{\heavy}}},\;0_{\mathfrak{n}_{\mathrm{\heavy}}},\; \overbar{\overbar{\kappa_{\elec}}}\cdot\nabla k_B T_{\elec},\;0_{3} \right)\nonumber\\
    \mathrm{S}^{\text{conv}}\left(\mathrm{U},\nabla \mathrm{U}\right) &=&\Biggl(0_{\mathfrak{n}_{\mathrm{\heavy}}},\;\left(n_\mathrm{\specie\ions} q_\mathrm{\specie\ions} \left( \left[\velocity_\mathrm{\specie\ions}+\velocity_{\elec}\right] \pvec \BField -\frac{\nabla P_{\elec}}{n_{\elec} q_{\elec}} \right)\right)_{\specie\ions\in\heavy},\nonumber\\ &&\,\quad\quad\quad\quad\quad\quad\quad\quad\quad\quad\quad\quad\quad-\left(P_\mathrm{\specie\ions}\nabla\cdot\velocity_\mathrm{\specie\ions}\right)_{\specie\ions\in\heavy},\;-P_\elec\nabla\cdot\velocity_\elec,\;-\nabla\pvec\left(\frac{1}{n_{\elec}q_{\elec}}\nabla P_{\elec}\right) \biggr)\nonumber\\
    \mathrm{S}^{\text{stiff}}\left(\mathrm{U}\right) &=&\biggl(\left(m_{\specie\ions}\Gamma^{\text{ion}}_{\mathrm{\specie\ions}}+ m_{\specie\ions}\Gamma^{\text{rec}}_{\mathrm{\specie\ions}}\right)_{\specie\ions\in\heavy},\;\left(\momex^{\text{ion}}_\mathrm{\specie\ions} +\momex^{\text{rec}}_\mathrm{\specie\ions} +\momex_\mathrm{\specie\ions}^{\text{col}} +\left[\frac{n_{\specie\ions}q_{\specie\ions}}{n_{\elec}q_{\elec}}\right]\momex_\mathrm{\elec}^{\text{col}}\right)_{\specie\ions\in\heavy},\; \nonumber\\ &&\,\quad\quad\quad\quad\quad\quad\quad\quad\quad\quad\quad\quad\quad\left(Q^{\text{ion}}_\mathrm{\specie\ions} + Q^{\text{rec}}_\mathrm{\specie\ions} +   Q_\mathrm{\specie\ions}^{\text{col}}\right)_{\specie\ions\in\heavy}, \;Q^{\text{ion}}_\mathrm{\elec} + Q^{\text{rec}}_\mathrm{\elec} +   Q_\mathrm{\elec}^{\text{col}},\;\nabla\pvec\left(\frac{\momex_\mathrm{\elec}^{\text{col}}}{n_\elec q_\elec}\right)\biggr)\nonumber
\end{eqnarray}
\noindent where $\mathrm{U}$ are the conservative variables of the system, $\mathrm{F}$ the convective fluxes, $\mathrm{D}$ the diffusive fluxes, $\overbar{\overbar{\kappa_{\elec}}}$ the anisotropic electron thermal conductivity,  $\mathrm{S}^{\text{conv}}$ the source terms as part of the convective system and $\mathrm{S}^{\text{stiff}}$ the stiff source terms associated with collisions and ionization/recombination processes, $0_{\mathfrak{n}_{\mathrm{\heavy}}}$ is a null vector of size $\mathfrak{n}_{\mathrm{\heavy}}$ and $\identity$ is the identity matrix.

Note that in this system the only diffusive flux is the one associated with the Spitzer term in the equation of thermal energy of the electrons. The source terms $\mathrm{S}^{\text{conv}}$  have characteristic timescales which typically can be much larger than $\mathrm{S}^{\text{stiff}}$. Indeed, $\mathrm{S}^{\text{stiff}}$ have timescales at the order of the collisional, ionization and recombination rates between all the particles $\specie\ions\in\mixture$. One can notice that the Lorentz force terms in momentum equations have been split into two parts, one in the $\mathrm{S}^{\text{conv}}$ part and the other one in $\mathrm{S}^{\text{stiff}}$. This is because the part of the Lorentz force as $\left[\frac{n_{\specie\ions}q_{\specie\ions}}{n_{\elec}q_{\elec}}\right]\momex_\mathrm{\elec}^{\text{col}}$ has a characteristic timescale which is at the order of the collisional time between electrons and heavy-particles, which can be much smaller than the other terms of the Lorentz force.

\bibliographystyle{aa}
\bibliography{aamnemonic,collectionbib}

\end{document}